\documentclass[aps,prl,twocolumn,groupedaddress,showpacs]{revtex4}
\usepackage{graphicx}

\begin{document}

%\preprint{}


\title{Finding the stable structures of W$_x$N$_{1-x}$ with an {\em ab-initio}  high-throughput approach}


\author{Michael J. Mehl}
\email[]{michael.mehl@nrl.navy.mil}
%\thanks{}
\affiliation{Center for Computational Materials Science, Naval  Research Laboratory, Washington DC 20375}
\author{Daniel Finkenstadt}
\author{Christian Dane}

\affiliation{Department of Physics, U.S. Naval Academy, Annapolis MD  21402}

\author{Gus L. W. Hart}
\affiliation{Department of Physics and Astronomy, Brigham Young University, Provo, UT 84602}
\author{Stefano Curtarolo}
\email[]{stefano@duke.edu}
\affiliation{Materials Science, Electrical Engineering, Physics and Chemistry, Duke University, Durham NC 27708}

\date{\today}

\begin{abstract}
Using density functional theory calculations, many researchers have
predicted that various tungsten-nitride compounds
WN$_x (x > 1)$ will be ``ultra-incompressible'' or ``superhard'',
{\em i.e.} as hard as or harder than diamond. These compounds are
predicted to have large bulk and shear moduli, ($>~200$~GPa) and to
be elastically and vibrationally stable. 

Compounds with such desirable properties must be energetically
stable against decomposition into other compounds. This stability
can only be found after the determination of the convex hull for
W$_x$N$_{1-x}$ lines which connect the lowest enthalpy structures as
a function of composition. The phase diagram of the W-N structure is
uncertain, both experimentally and computationally. Complex van der
Waals forces play a significant role in determining the structure of
solid N$_2$.

Here we use high-throughput calculations to map out the
convex hull and other low energy structures for the W-N system.  We
find that the ground state of the system is the NbO structure, and
that the WN$_2$ structures found by Wang {\em et al.} are also
stable when van der Waals forces are neglected. Other proposed
structures are above the convex hull of the W-N system. We show
how the choice of density functional influences the shape of the
curve and the structures that form the hull. In nitrogen-rich
compounds, the choice of functional can dramatically change the
structural parameters and mechanical behavior. Using any of the
functionals, the bulk and shear moduli of the NbO phase are
comparable to the WN$_x$ compounds that have been claimed to be
ultra-incompressible or superhard.
\end{abstract}

%\keywords{}

\maketitle

\section{Introduction \label{sec:intro}}

The search for materials with hardness comparable to diamond has
centered on compounds with highly directional, short, and strong
bonds.\cite{kaner05:hardmat} In addition to diamond, these include
other proposed carbon
structures,\cite{ribeiro06:hypocarb,lyakhov11:hardmat} cubic boron
nitride\cite{Mousang63:BN}, carbonitrides,\cite{veprek99:hardmat}
and transition metal borides.\cite{ivanovskii12:hpmat,hao06:XB2}
Over the past several years there has been considerable theoretical
interest in tungsten nitride
systems.\cite{suetin08:wc-wn,wang09:wn2,song10:wn,aydina12:XN4}
These computational studies predict that some compounds WN$_x$ with
$x \ge 2$ have large bulk and shear moduli. In one
case\cite{aydina12:XN4} the hardness of the material was estimated
to be on the same order as boro- and carbo-nitrides.

The present studies all address the issue of stability for various
predicted WN$_x$ compounds. This is done by showing they are elastically
and/or vibrationally stable. It is also ensured the formation
enthalpy of the compound is lower than the formation energy of its
components, bcc tungsten, and $\alpha$N$_2$.\cite{donohue74:elements}
That is,
\begin{widetext}
  \begin{equation}
    \label{equ:enthalpy}
    \Delta H (\mbox{W}_x \mbox{N}_y) = [E (\mbox{W}_x \mbox{N}_y) - x
      \, E (\mbox{W}) - (y/2) \, E(\alpha \mbox{N}_2) ] /(x + y) ~ .
  \end{equation}
\end{widetext}
where $\Delta H (\mbox{W}_x \mbox{N}_y)$ is the formation energy per
atom of a compound with stoichiometry W$_x$N$_y$, $E (\mbox{W}_x
\mbox{N}_y)$ is the energy/formula unit of the compound,
$E(\mbox{W})$ is the equilibrium energy per atom of body-centered
cubic tungsten, and $E(\alpha \mbox{N}_2)$ is the equilibrium energy
per molecule of solid $\alpha$N$_2$.

Condition (\ref{equ:enthalpy}) is necessary, but not sufficient, for
structural stability. To be truly stable, a structure must lie on
the convex hull constructed from the plot of enthalpy/atom versus
composition for all possible phases of the system. As we shall see,
most of the WN$_x$ structures referred to above do not fulfill this
criterion.

The construction of the convex hull for the tungsten nitride
system is non-trivial. The experimental phase diagram is not well
defined.\cite{wriedt89:WNphases} Various papers have presented
experimental results for
cubic\cite{hagg30:WN,khitrova59:WN,khitrova62:WN,khitrova62:WN2}
(``$\beta$'', composition 33-50\% Nitrogen), hexagonal
(``$\delta$'', composition 33-67\% Nitrogen)\cite{khitrova62:WN},
and tungsten-carbide (WC)\cite{schonberg54:MoNWN} structures. Many
of these are only found in thin films. There is no clear preference
for a ground state structure at any composition. As a result, there
is little guidance for computational first-principles studies of
W-N. The computations that do exist start from a wide variety of
structures.

Using an evolutionary technique, Wang {\em et al.},\cite{wang09:wn2}
found that the lowest energy structure of WN$_2$ is hexagonal. The
compound has either an hP3 structure, with space group
$P\overline{6}m2$, or an hP6 structure with space group $P6_3/mmc$.
At other stoichiometries the calculations are less certain.
Previous papers have considered the ground state of WN to be
NaCl,\cite{isaev07:tmcn}, tungsten carbide
(WC),\cite{suetin08:wc-wn,song10:wn} or NiAs.\cite{kroll05:TaN_WN}
More recently, Song {\em et al.} \cite{song10:wn} looked at WN$_3$
in the P$_3$Tc structure, while Aydin {\em et
  al.}\cite{aydina12:XN4} have suggested that WN$_4$ has the ReP$_4$
structure.

None of these calculations addressed the complete range of
stoichiometries in the WN system.  What is needed is a method which
quickly examines a wide variety of possible structures over large
ranges of composition. One such method is
{\small AFLOW},\cite{curtarolo:art65} a high-throughput front end for
electronic structure calculations.\cite{curtarolo:art81} {\small AFLOW}
allows us to quickly examine a selected range of structures using
high-performance supercomputers and modern density functional
electronic structure techniques. The {\small AFLOW} prototypes' database, which was
originally used to describe intermetallic alloys, \cite{curtarolo:art57,curtarolo:art63,curtarolo:art67,curtarolo:art70,curtarolo:art87} can easily be
enlarged.\cite{aflowlibPAPER} We were therefore able to include
ionic and covalent structures which seem chemically similar to
W-N. These include borides, carbides, oxides, and other nitrides.

In this paper we examine the possible ground states of the W-N
system over a wide range of stoichiometries. Using the power of
{\small AFLOW}, we can examine hundreds of possible W$_x$N$_y$ structures. We
show that the true ground state of W-N is related to what is known
as $\beta$WN, and show why it exists over a wide range of
stoichiometries. In addition, the bulk and shear moduli found for
$\beta$WN are comparable to those WN$_x$ compounds predicted to be
ultra-incompressible\cite{wang09:wn2} or
superhard.\cite{aydina12:XN4}

The paper is organized as follows: Section~\ref{sec:methods}
describes the computational methods used. Section~\ref{sec:hull}
describes the main calculations of this paper: the determination of
the convex hull for WN as a function of enthalpy
(\ref{equ:enthalpy}). The majority of the calculations are done
using the Perdew-Burke-Ernzerhof (PBE)\cite{perdew96:pbegga}
implementation of a Generalized Gradient Approximation density
functional.

The ground state $\alpha$N$_2$ structure of Nitrogen is influenced
by van der Waals forces. We investigate these non-covalent forces
using different density functionals in
Section~\ref{sec:otherdft}. In particular, the Local Density
Approximation (LDA) \cite{ceperley80:egas,perdew81:_sic}
calculations are presented in Section~\ref{sec:ldacomp}, and the
so-called DF2 van der Waals functional (vdW-DF2)\cite{dion04:vdwdft}
is discussed in Section~\ref{sec:vdwcomp}.

We then discuss several specific structures:
Section~\ref{sec:betanw} we examine the cubic $\beta$WN phase, and
the competing $\delta$WN phase in
Section~\ref{sec:deltanw}.\cite{khitrova62:WN} Section~\ref{sec:WN2}
looks at the WN$_2$ structures found by Wang {\em et
  al.},\cite{wang09:wn2} along with other low energy WN$_2$
structures.

Section~\ref{sec:WN3} studies possible WN$_3$ structures, including
the P$_3$Tc structure proposed by Song {\em et al.},\cite{song10:wn}.
Section~\ref{sec:WN4} looks at the ReP$_4$ structure of WN$_4$
predicted by Aydin and coworkers.

We conclude with Section~\ref{sec:discuss}, which summarizes the
results, and discuss the effects of the choice of density functional
on the predictions for this system.

Finally, as a complete reconstruction of a the unit cell is
necessary for the reproduction of any calculation, we list the
crystallographic information for the most important structures
discussed in this paper in the supplementary material.

\section{Methods \label{sec:methods}}

We began our search for the convex hull of the W-N system by using
{\small AFLOW} \cite{curtarolo:art65,curtarolo:art81,curtarolo10:aflow} to
quickly and efficiently search through a large database of
structures. As the original {\small AFLOW} prototypes' database was for binary metallic
alloys, \cite{curtarolo:art57,curtarolo:art63,curtarolo:art67,curtarolo:art70,curtarolo:art87} 
we extended it to include over fifty new structures. These
include nitrides, oxides, borides, and carbides. The important
structures, including all of those discussed in this paper, are
described in the supplementary material.

Electronic structure calculations were done using the Vienna {\em Ab initio} Simulation Package
(VASP),\cite{kresse93:vasp1,kresse94:vasp2,kresse93:vasp3} including
core state effects via the VASP
implementation\cite{kresse99:ustopaw} of the Projector
Augmented-Wave (PAW) method.\cite{blochl94:paw} Since some our
results were unexpected, we checked them against computations
performed with ELK, an all-electron full-potential Linearized
Augmented Plane Wave (FP-LAPW) code.\cite{dewhurst:elk}

{\small AFLOW}'s default is to use the Perdew-Burke-Ernzerhof (PBE)
implementation\cite{perdew96:pbegga} of the Generalized Gradient
Approximation (GGA) to Density Functional Theory
(DFT).\cite{hohenberg64:dft,kohn65:inhom_elec} There are, however,
substantial differences between PBE results and those from the Local
Density Approximation (LDA).\cite{ceperley80:egas,perdew81:_sic}
This was shown by Wang {\em et al.}\cite{wang09:wn2}, and we shall
see further examples below. Accordingly, we have done some
calculations using the LDA implemented in VASP, as shown in
Section~\ref{sec:ldacomp}.

Molecular nitrogen, $\alpha$N$_2$, is a van der Waals
solid,\cite{eremets01:nitrogen} and neither the LDA nor the PBE-GGA
correctly predict its lattice constant. We studied the effect of van
der Waals forces by utilizing the VASP
implementation\cite{klimes10:vdwdft,klimes11:vdWtest} of the vdW-DF2
functional.\cite{dion04:vdwdft} This functional was developed to
handle the van der Waals interaction in general geometries, so we
apply it over the entire range of compositions. These results are
discussed in Section~\ref{sec:vdwcomp}.

We used the VASP-supplied LDA and PBE PAW potentials for Nitrogen
and Tungsten (specifically, May 2000 LDA and April 2002 PBE PAW
potentials for N, and the July 1998 LDA and September 2000 Tungsten
W\_pv PAWs) from the VASP distribution. All calculations use a
kinetic energy cutoff of 560~eV, which is 40\% larger than the
suggested cutoff for Nitrogen (400~eV). Unless otherwise stated, we
let the {\small AFLOW} package determine the k-point mesh for each
structure. This is usually set to use approximately the same density
of k-points in reciprocal space for all structures.

For the NaCl structure, {\em
  e.g}, a $15 \times 15 \times 15$ Monkhorst-Pack
grid\cite{monkhorst76:special_k} was originally used. For Wang
{\em et al.}'s $P\overline{6}m2$ WN$_2$ structure we used a $15
\times 15 \times 9$ mesh. In the initial VASP calculations performed
via {\small AFLOW} all structures are considered to be spin-polarized. However, in
every structure the self-consistent magnetic moment was negligible,
and the final calculations were all done assuming no moment.

\begin{figure*}
  \includegraphics[width=17cm]{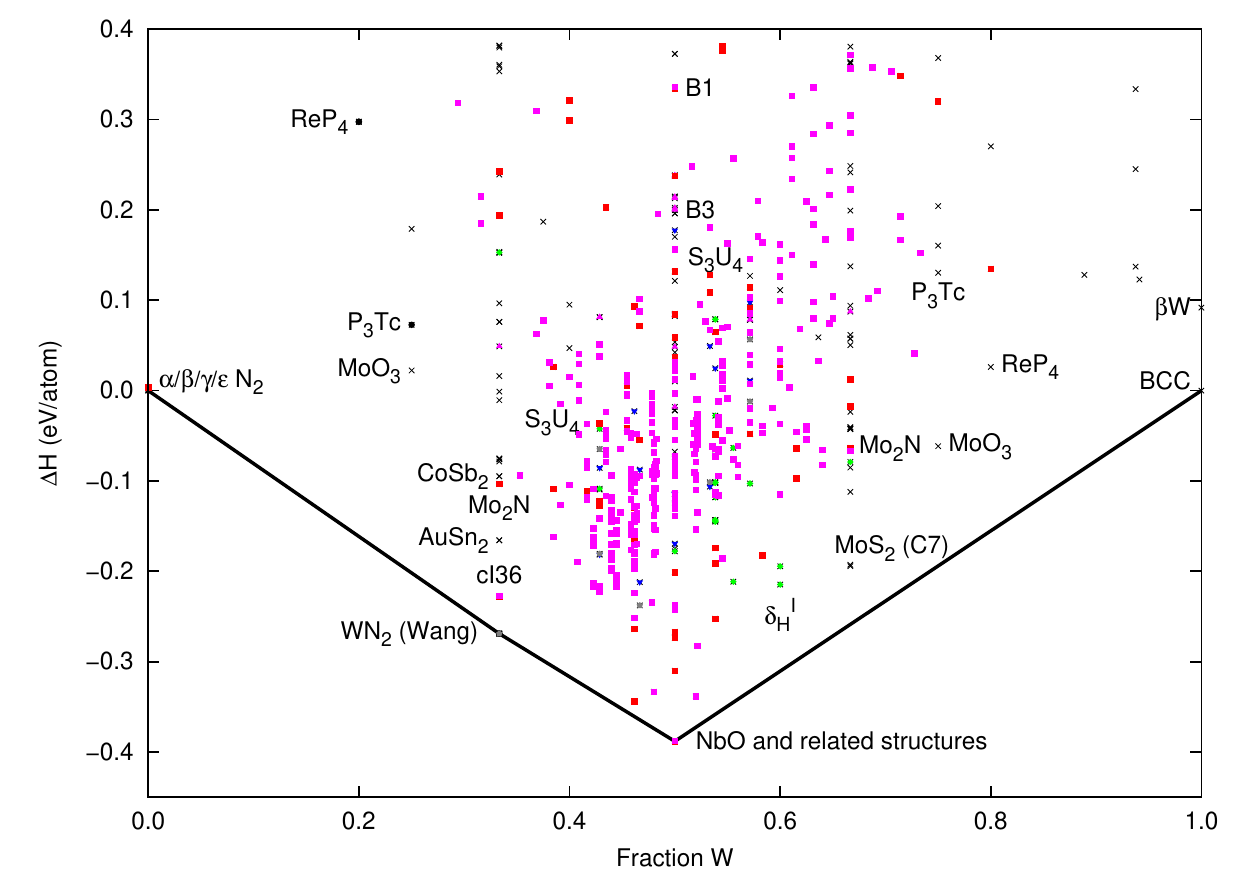} % {equil_x1}
  \caption{\small (Color online) Low-enthalpy structures of Tungsten
    Nitride, calculated using PBE. Each cross gives the enthalpy of
    a structure from the {\small AFLOW} database, as described in the
    supplementary materials. The solid colored symbols denote
    structures constructed by removing atoms from a base structure
    and allowing the system to relax, as described in the text and
    the supplementary material. Red squares: NaCl and constructed by
    removing atoms from supercells of NaCl, including possible
    $\beta$-phase structures.\protect\cite{hagg30:WN} Green
    diamonds: the $\delta$-phase and related
    structures.\protect\cite{khitrova62:WN} Upward pointing purple
    triangles: Cubic ZnS (B3) and related structures. Downward
    pointing blue triangles:  NiAs (B8$_1$). Gray pentagons: WC and
    related structures. The black circles indicate structures which
    other researchers have described as
    hard.\protect\cite{wang09:wn2,aydina12:XN4} Note that Wang {\em
      et al.}'s $P\overline{6}m2$ WN$_2$ structure could be grouped
    with the NiAs structures as well as the WC structure.
    \label{fig:hull}}
\end{figure*}

The MedeA software system\cite{medea} was used to drive VASP in
calculations of the phonon spectra and some other calculations.
Elastic constants were computed by taking finite
strains\cite{mehl90:_struc,mehl95:fpcij} and determining the slope
of the corresponding stress-strain curve (as implemented in the
MedeA package, or using the native VASP option). Phonon frequencies
were determined via the frozen-phonon approximation using the MedeA
package.

\section{The Ground State Hull and low-lying states of Tungsten
  Nitride \label{sec:hull}}

Overall we considered 350 possible structures from the expanded
{\small AFLOW} prototypes' database. The equilibrium configurations for the important
structures are described in the supplementary material.
Fig.~\ref{fig:hull} shows the formation enthalpy per atom compared
to $\alpha$N$_2$ and bcc-W, as defined in (\ref{equ:enthalpy}).

\begin{figure}[htb]
  \includegraphics[width=7cm]{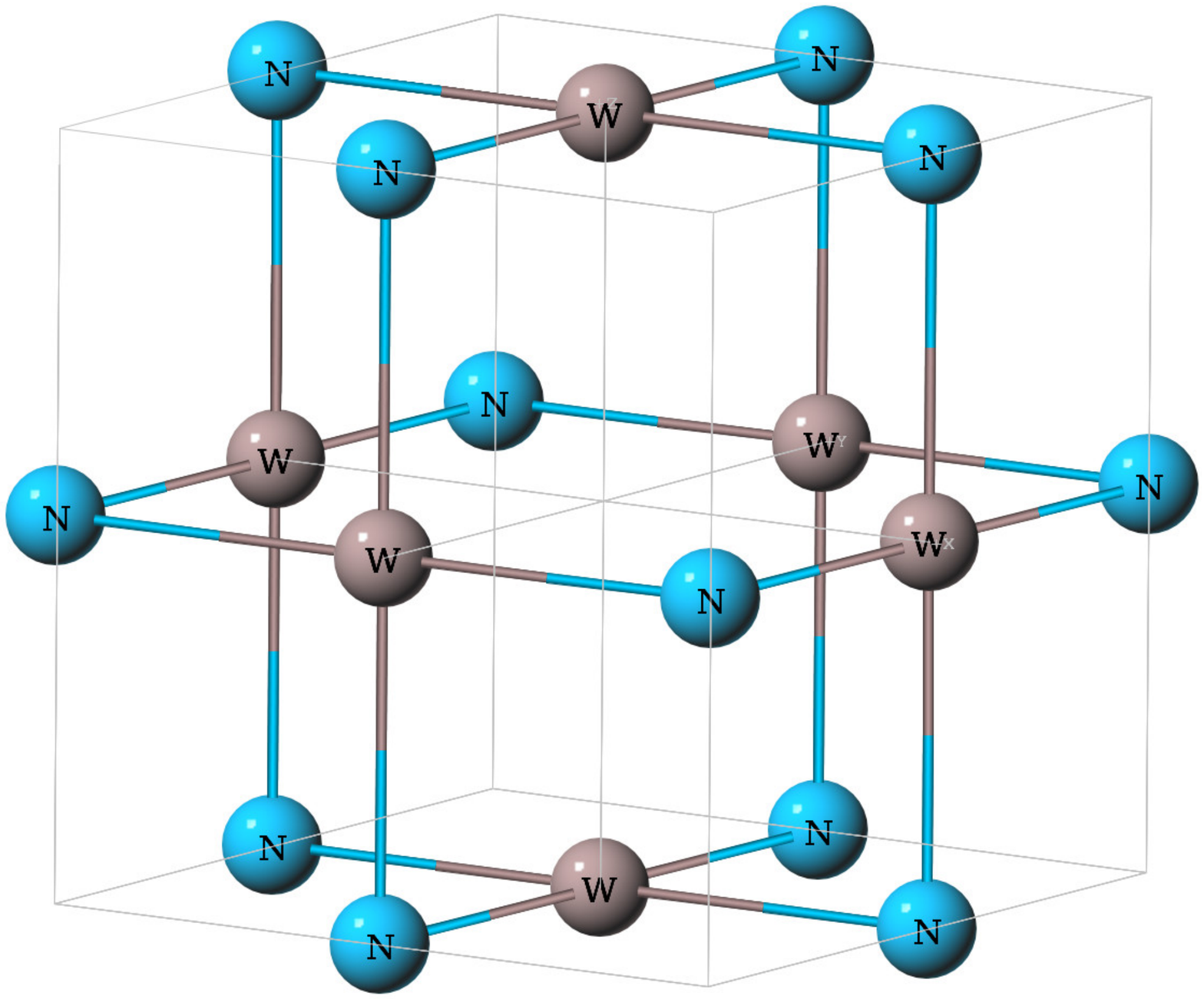} %{NbO}  
  \vspace{-5mm}
  \caption{\small \label{fig:NbO} (Color Online) WN in the NbO structure,
    constructed by taking equally spaced vacancies from both sites of
    the NaCl structure. The PBE functional predicts the equilibrium
    lattice constant to be 4.131\AA.}
\end{figure}

The other structures on the convex hull in Fig.~\ref{fig:hull} are
the two WN$_2$ structures found by Wang {\em et
  al.}\cite{wang09:wn2}, and the NbO structure (Fig.~\ref{fig:NbO}),
which will be considered further in Section~\ref{sec:betanw}. To our
knowledge this structure has never been considered as a candidate
structure for WN. Its existence is not a complete surprise: WN has
the same number of valence electrons as NbO, and the Nb and W atoms
have similar valence shells ($4d^4 5s$ for Nb, $5d^4 6s^2$ for
W). The electron gained by replacing Nb with W is lost again as we
replace O with N.

The NbO structure is a supercell of the NaCl structure, with ordered
vacancies on both the Na and Cl sites. Many of the other low-lying
structures shown in Fig.~\ref{fig:hull} can be constructed by
removing atoms from a supercell of the parent structure. Starting
with the NaCl (B1), cubic ZnS (B3), and NiAs (B8$_1$) structures, we
constructed supercells of the original cell (as described below) and
removed atoms from both sites in a systematic manner. These
structures are color/shape coded in Fig.~\ref{fig:hull}. In all
three cases, some supercells with multiple vacancies had lower
energy than the parent structure. The ground state NbO can be
constructed by removing atoms from the parent NaCl compound (NbO),
while the $P\overline{6}m2$ WN$_2$ structure can be derived from
either the NiAs or the tungsten carbide structure.

Some structures which have been predicted to be stable actually have
enthalpies far from the convex hull. In particular, NaCl is over
0.3~eV/atom above NbO, and the WC and NiAs structures are over
0.2~eV/atom above NbO. We will discuss the predicted P$_3$Tc
structure for WN$_3$ and ReP$_4$ structure for WN$_4$ in later
sections.

\section{Results using different density
  functionals \label{sec:otherdft}}

As noted above, solid $\alpha$N$_2$ is bound by the van der Waals
interaction between molecules. Furthermore, Wang~{\em et al.} found
a substantial change in the enthalpy of their WN$_2$ systems when
changing from PBE to LDA functionals. This suggests that the choice
of functional may well change our predictions of the structure in
the W-N system. We tested this by comparing results for selected
structures using three different density functionals: the
Perdew-Burke-Ernzerhof Generalized Gradient Expansion
(PBE),\cite{perdew96:pbegga} the Local Density Approximation
(LDA),\cite{ceperley80:egas,perdew81:_sic} and the van der Waals
functional developed by Dion {\em et al.}
(vdWDF2),\cite{dion04:vdwdft} as implemented in VASP by Klime\v{s}
{\em et al.}\cite{klimes10:vdwdft,klimes11:vdWtest} We began our
tests by considering the known ground states of the end-points,
$\alpha$N$_2$ and bcc-W.

First consider nitrogen. There is a computational problem in using
VASP to compute the equilibrium lattice constant for all of the
nitrogen structures considered here. The bulk modulus of the
molecular N$_2$ crystal in any phase is extremely small, about
1~GPa.\cite{kittel96:issp} As a result, algorithms which stop
searching for a minimum when the pressure reaches some small number
will fail here unless an extremely small tolerance is used. In
addition, while the calculation of the total energy of a system is
variational, that of the pressure is not. As a result, the pressure
calculation will not converge unless we use a much larger plane-wave
basis set. We eliminated this problem for the N$_2$ phases by
explicitly calculating $E(V)$ at discrete points, bounding the
minimum, and using a fourth-order Birch fit\cite{birch78:eos} to
determine the equilibrium lattice constant.

Experiments have determined that $\alpha$N$_2$ is cubic with either
space group $P2_13$-$T^4$ (\#195), which has no inversion site, or
$Pa\overline{3}$-$T_h^6$ (\#205), which contains an
inversion.\cite{donohue74:elements} In VASP the energy and
structural differences between the two structures is negligible. In
fact, the energy difference between the five N$_2$ structures shown
in Fig.~\ref{fig:hull} is less than 4~meV/atom.  This being the
case, here we present only the results for the highest symmetry
structure, $Pa\overline{3}$. The results for the equilibrium
geometry of this phase for all three functionals are given in
Table~\ref{tab:N2}.

\begin{table}
  \caption{\small Equilibrium lattice constant, internal parameter, and
    equilibrium bulk modulus for the $Pa\overline{3}$ structure of
    $\alpha$N$_2$, as determined using various density functionals
    and compared to
    experiment.\cite{donohue74:elements,kittel96:issp} The lattice
    is simple cubic, and the nitrogen atoms sit on the (8c) Wyckoff
    position, which has one internal parameter, $x$.  The
    equilibrium lattice constant and bulk modulus are determined
    from a fourth order Birch fit. The quantity d(N-N) is the length
    of the nitrogen-nitrogen bond in N$_2$
    molecules. \label{tab:N2}}
  \begin{tabular}{{c|c|c|c|c}}
    \hline\hline
    Functional & PBE & LDA & vdW & Exp\cite{donohue74:elements} \\
    \hline
    a (\AA) & 6.187 & 5.223 & 5.511 & 5.659 \\
    $x$ & 0.052 & 0.061 & 0.058 & 0.056 \\
    d(N-N) (\AA) & 1.11 & 1.10 & 1.11 & 1.10 \\
    K$_0$ (GPa) & 0.788 & 5.70 & 4.69 & 1.2\cite{kittel96:issp} \\
  \end{tabular}
\end{table}

\begin{table}
  \caption{\small Equilibrium lattice constant and bulk modulus for bcc
    tungsten, as determined using various density functionals and
    compared to experiment.\cite{donohue74:elements,simmons71:cij}
    The equilibrium lattice constant and bulk modulus are computed
    from a fourth order Birch fit to energy versus volume
    data. \label{tab:bcc-W}}
  \begin{tabular}{{c|c|c|c|c}}
    \hline\hline
    Functional & PBE & LDA & vdW & Exp \\
    \hline
    a (\AA) & 3.190 & 3.143 & 3.250 & 3.165\cite{donohue74:elements} \\
    K$_0$ (GPa) & 304 & 337 & 267 & 323\cite{simmons71:cij} \\
  \end{tabular}
\end{table}

The PBE GGA overestimates the equilibrium lattice constant of
$\alpha$N$_2$ by 9\% while the LDA underestimates it by almost 8\%,
as was also found by Mailhoit {\em et
  al.}.\cite{mailhiot92:nitrogen} Both of these errors are much
larger than found for most elements.\cite{perdew92:gga-apps} If we
consider the van der Waals interaction using the vdW-DF2 functional,
we get an error in the lattice constant of 5.2\%, which, while not
ideal, is a substantial improvement on both the LDA and PBE
results. All of the calculations give essentially the same distance
for the N-N bond, suggesting that errors in the calculations are due
to the long-range interaction between the molecules. Both LDA and
vdW-DF2 substantially overestimate the bulk modulus, while PBE
underestimates it by about 35\%. The vdW-DF2 prediction demonstrates
the importance of van der Waals interactions in Nitrogen systems.

Similar calculations were carried out for the ground state
body-centered cubic structure of tungsten. The results are presented
in Table~\ref{tab:bcc-W}. In this case the equilibrium lattice
constant determined by VASP is almost exactly identical to the
result from a Birch fit to energy-volume data. We use the latter
so that we can also compute the equilibrium bulk modulus and
compare it to experiment.\cite{simmons71:cij} All three functionals
produce reasonable values for the equilibrium lattice constant and
bulk modulus, with the LDA yielding perhaps the best agreement.

The conclusion of this brief study is that none of the three
functionals is ideal for describing both tungsten and nitrogen
structures. All three do give the same ordering of the low-energy
states at both endpoints. As we shall see below, this agreement does
not extend to W$_x$N$_{1-x}$ compounds, and the choice of functional
significantly affects the predicted shape and structural composition
of the convex hull.

\subsection{LDA results \label{sec:ldacomp}}

\begin{figure}
  \includegraphics[width=8.5cm]{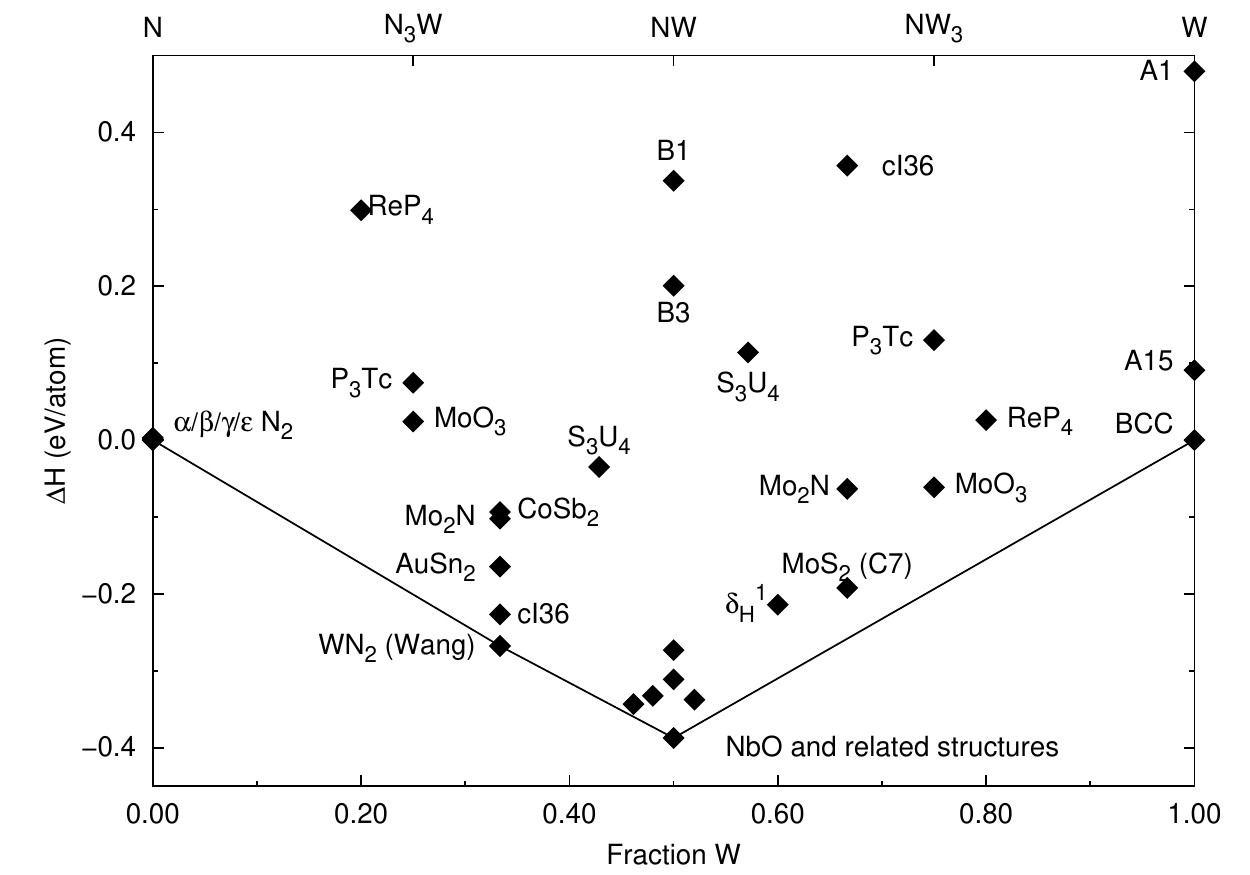} % pbe}
  \caption{\small The relative enthalpy (\ref{equ:enthalpy}) of structures
    making up the ground state hull. and other selected structures
    for the Tungsten Nitride system. These were calculated using the
    Perdew-Burke-Ernzerhof (PBE) functional\cite{perdew96:pbegga}
    with VASP. The state labeled cI36 at x~=~1/3 is the NbO
    structure with additional defects, as described in
    Section~\ref{sec:betanw}. \label{fig:pbehull}}
\end{figure}

\begin{figure}
  \includegraphics[width=8.5cm]{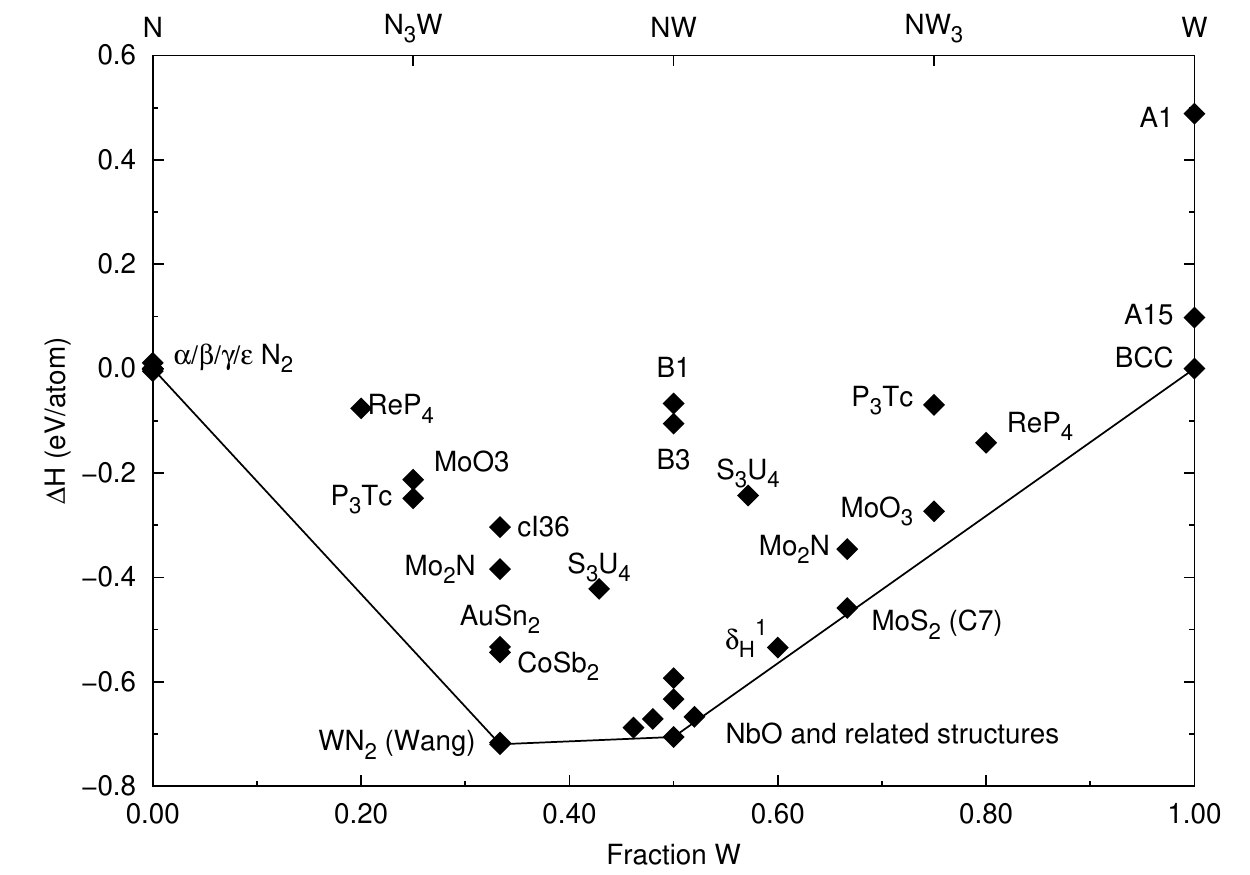} %lda}
  \caption{\small The relative enthalpy (\ref{equ:enthalpy}) of structures
    making up the ground state hull, and other other structures
    for the Tungsten Nitride system. These were calculated using the Local
    Density Approximation (LDA) with VASP. The state labeled cI36 at
    x~=~1/3 is the NbO structure with additional defects, as described
    in Section~\ref{sec:betanw}. \label{fig:ldahull}}
\end{figure}

The calculations leading to the results in Table~\ref{tab:N2} show
that the PBE functional improperly describes the ground state
structure of $\alpha$N$_2$. This calls into question the predicted
shape of the ground state hull shown in Fig.~\ref{fig:hull}.  We
examine the possible errors in the PBE by redoing a subset of our
calculations with other density functionals. We have selected a
data-set of 34 structures, including end points, low-lying states
(WN$_2$, NbO and related structures, {\em etc.}), ``interesting''
states (P$_3$Tc, ReP$_4$), and ``parent'' structures (NaCl,
B3-ZnS). All these are described in the supplementary material. The
resulting enthalpy calculation is shown in Fig.~\ref{fig:pbehull}
for the PBE functional, and Fig.~\ref{fig:ldahull} for LDA.

The composition of the hull is unchanged when going from PBE to
LDA. Its shape changes visibly, as the hexagonal WN$_2$ phases are
now competitive with the NbO phase. Within the LDA, all structures
are more tightly bound compared to the endpoints, by a factor of two
or more. On the tungsten right side of the LDA plot two structures,
$\delta_H^I$ (discussed in Section~\ref{sec:deltanw}) and MoS$_2$,
are extremely close to the tie-line between the NbO structure and
bcc tungsten. This indicates that the LDA functional may well be the
best choice for calculations on the tungsten-rich side, since
several of $\delta$ phases are observed
experimentally.\cite{khitrova62:WN}

\subsection{van der Waals results \label{sec:vdwcomp}}

\begin{figure}
  \includegraphics[width=8.5cm]{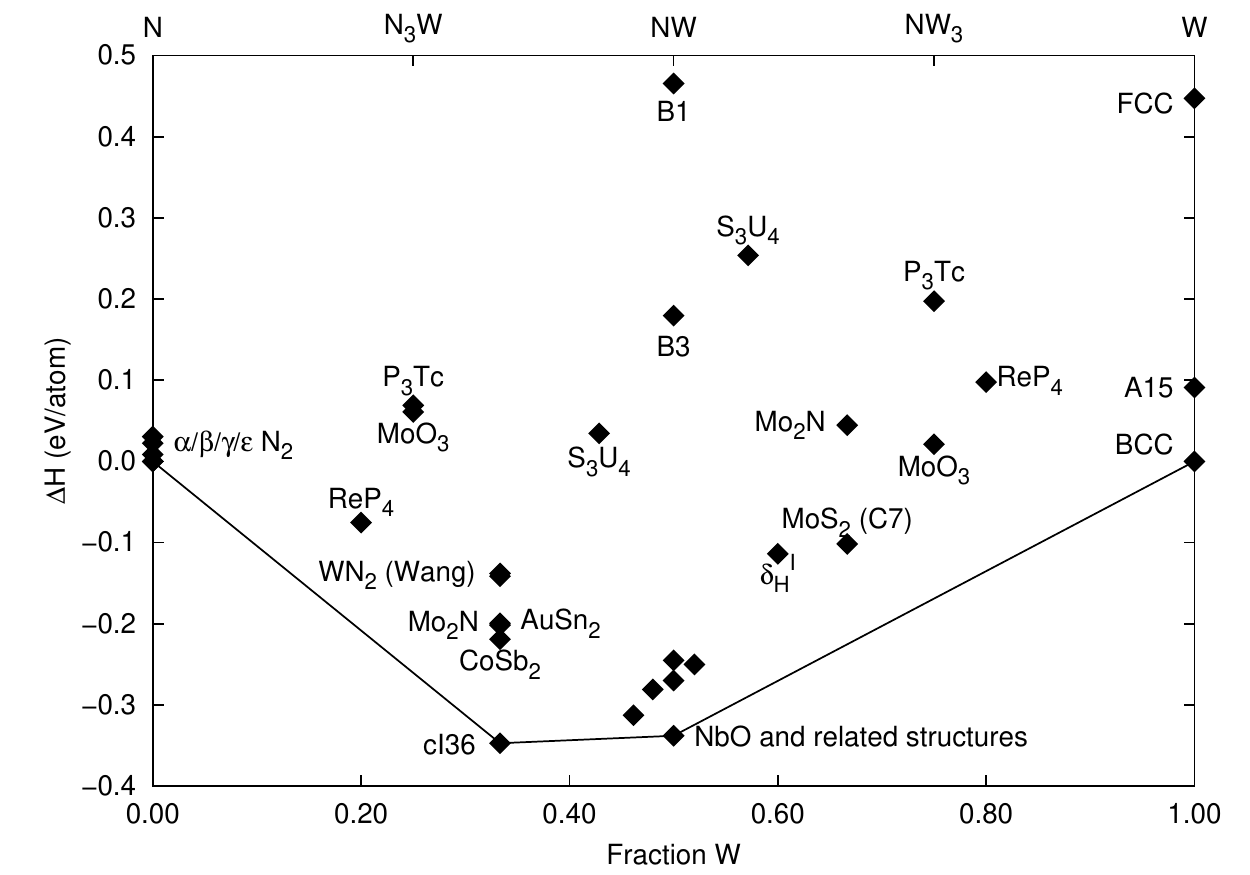} %vdW}
  \caption{\small The relative enthalpy (\ref{equ:enthalpy}) of structures
    making up the ground state hull and selected other structures
    for the Tungsten Nitride system. These were calculated using
    vdW-DF2 van der Waals functional with VASP. The state labeled
    cI36 at x~=~1/3 is the NbO structure with additional defects, as
    described in the Section~\ref{sec:betanw}. \label{fig:vdwhull}}
\end{figure}

At first glance, Fig.~\ref{fig:vdwhull}, which shows the vdW-DFT
convex hull for W$_x$N$_{1-x}$, looks similar to the PBE
(Fig.~\ref{fig:hull}) and LDA (Fig.~\ref{fig:ldahull}) diagrams. On
closer examination we see substantial differences. The principle
difference is that the vertex of the hull at x~=~1/3 is not one of
the WN$_2$ structures found by Wang {\em et al.}. Rather, it is a
cubic structure we have labeled cI36. In fact, the Wang structures
are above a line drawn from $\alpha$N$_2$ and NbO, and so would not
be on the hull even if the cI36 structure was not present.  The
ReP$_4$ structure is now rather low in energy, about 0.1~eV/atom
away from the tie line. We will examine these structures in more
detail in later sections.

\section{Cubic ($\beta$) WN structures \label{sec:betanw}}

H\"{a}gg\cite{hagg30:WN,schonberg54:MoNWN} identified the $\beta$
phase in the WN system as cubic, with the approximate composition
W$_2$N. The tungsten atoms form a face-centered cubic lattice and
the nitrogen atoms are randomly distributed, presumably at the
octahedral sites.\cite{khitrova62:WN} Chiu and
Chuang\cite{chiu93:WNfilms} grew thin films of nominal composition
WN using metallo-organic chemical vapor deposition (MOCVD). They
found a cubic structure of composition WN$_x$, with $0.7 < x < 1.8$.
Where the composition is near W$_2$N they found a lattice constant
of 4.125\AA, rising to 4.154\AA \, at stoichiometry and to 4.172\AA
\, as $x$ approaches 1.8. They state that the tungsten atoms remain
on FCC sites. For $x = 1$ the nitrogen atoms fill the octahedral
sites. For $x < 1$ there are vacancies on the octahedral sites, and
for $x > 1$ nitrogens populate both the octahedral and tetrahedral
sites.

Computationally, we find something different, as seen in
Figures~\ref{fig:pbehull}-\ref{fig:vdwhull}. All three functionals
predict NbO to be the ground state structure. NbO has space group
$Pm\overline{3}m$ (\#221). The tungsten atoms are on the (3c)
Wyckoff sites, and the nitrogen atoms are on the (3d) sites (or {\em visa
 versa}). It can be described as an eight-site supercell of the
sodium chloride structure with ordered vacancies on both the Na and
Cl sites, as shown in Fig.~\ref{fig:NbO}. Within the PBE-GGA, we
find the equilibrium lattice constant to be 4.131\AA. 

This is in reasonable agreement with the experimental value of
4.125\AA ~found by Chiu and Chaung\cite{chiu93:WNfilms} and reported
in the Alloy Phase Diagrams.\cite{wriedt89:WNphases} Chiu and Chaung
describe the site as having all FCC and octahedral sites filled,
{\em i.e.} the NaCl (B1) structure. Instead we find vacancies on
both sites. The minimization of the actual B1 structure gives a
lattice constant of 4.366\AA, in close agreement with other
calculations.\cite{isaev07:tmcn,suetin08:wc-wn} This suggests that
what Chiu and Chaung were seeing was actually the NbO structure, or
something very close to it.

Table~\ref{tab:nbo} lists the structural and elastic parameters of
WN in the NbO structure using all three functionals. The structure
is elastically stable, with all of the Born
criteria\cite{born40:stability} satisfied. We also estimate the
shear modulus using the average of the Hashin-Shtrikman
bounds.\cite{hashin62:var,hashin62:poly} The bulk and shear moduli
are comparable to those found by Wang {\em et al.}  for the
predicted hexagonal WN$_2$ structures, the WN$_3$ structure studied
by Song {\em et al.},\cite{song10:wn} and the WN$_4$ structure
predicted by Aydin {\em et al.}\cite{aydina12:XN4} We have not tried
to predict the actual hardness of the material, but the cubic NbO
structure of WN is certainly difficult to compress.

\begin{table}
  \caption{\small \label{tab:nbo} Equilibrium lattice and elastic constants
    of WN in the NbO structure, (Space Group $Pm\overline{3}m$
    \#221, Wyckoff positions (3c) and (3d)). These were computed by
    VASP using the appropriate PAW potentials for each
    exchange-correlation functional. Elastic constants (in GPa) were
    computed by finite strain
    calculations.\cite{mehl90:_struc,mehl95:fpcij} The isotropic
    shear modulus $G$ is the average of the Hashin-Shtrikman bounds
    for a cubic system.\cite{hashin62:var,hashin62:poly}}
  \begin{tabular}{lcccccc}
    \hline\hline
    Functional & a (\AA) & C$_{11}$ & C$_{12}$ & C$_{44}$ & B & G \\
    \hline
    LDA     & 4.078 & 884 & 140 & 177 & 388 & 239 \\
    PBE     & 4.131 & 754 & 126 & 173 & 335 & 220 \\
    vdW-DF2 & 4.208 & 655 & 120 & 154 & 298 & 192
  \end{tabular}
\end{table}

\begin{figure}
  \includegraphics[width=6.5cm,angle=270]{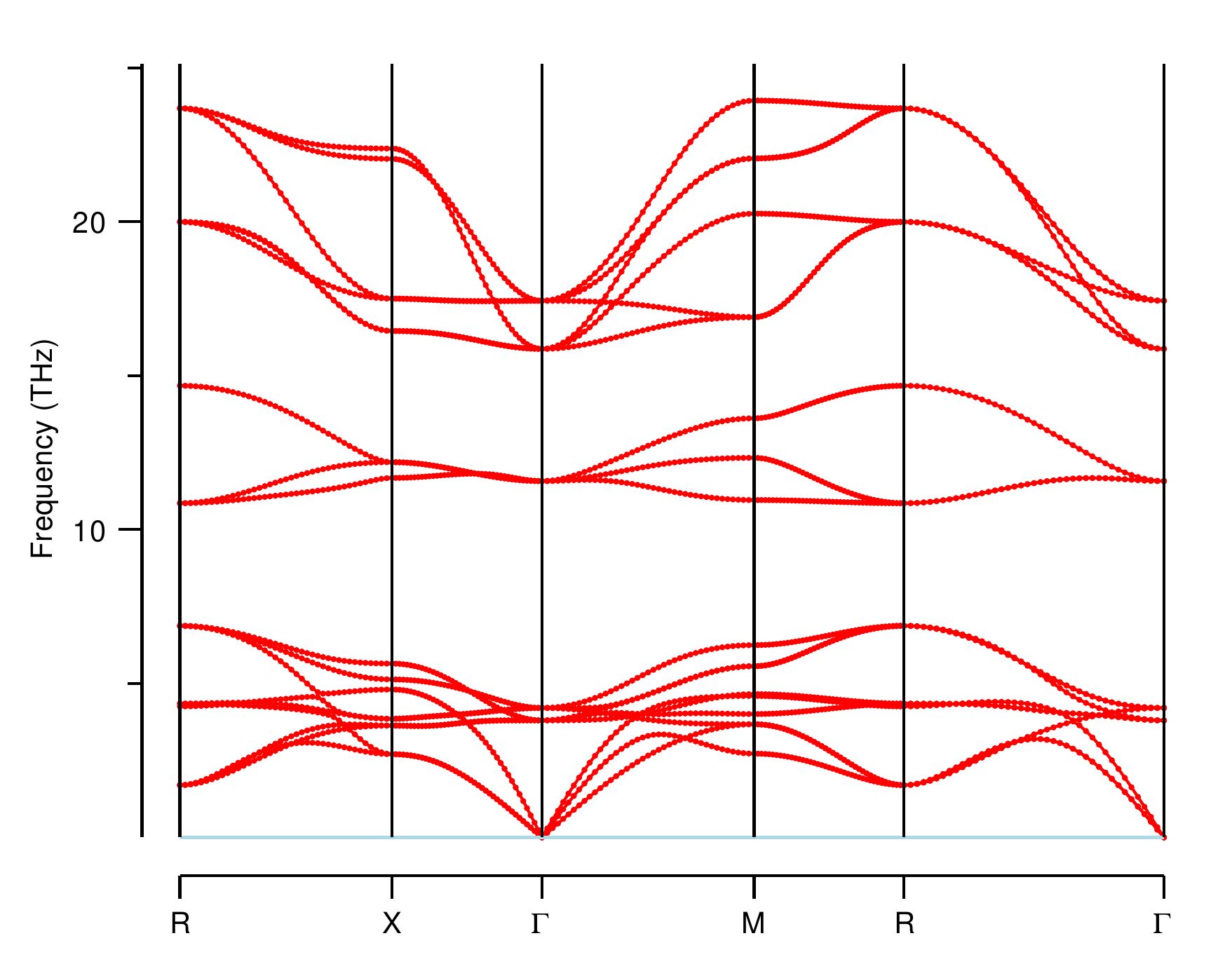} %NbOphonons}
  \caption{\small \label{fig:nboph} (Color Online) Phonon frequencies along
    high-symmetry lines for the NbO structure of WN, found via the
    MedeA package.\cite{medea} No imaginary phonon frequencies were
    found, indicating that this structure is stable against small
    amplitude vibrations.}
\end{figure}

We also checked the vibrational stability of the NbO structure by
computing the phonons along high symmetry lines using the MedeA
software package.\cite{medea} The results for the PBE calculation
are shown in Fig.~\ref{fig:nboph}. We found no imaginary frequencies
over the entire Brillouin zone, leading us to the conclusion that
the NbO structure of WN is at least metastable. Since this is the
lowest energy structure found, we believe that this is the stable
ground state structure of WN at this composition.

\begin{figure}
  \includegraphics[width=8.5cm]{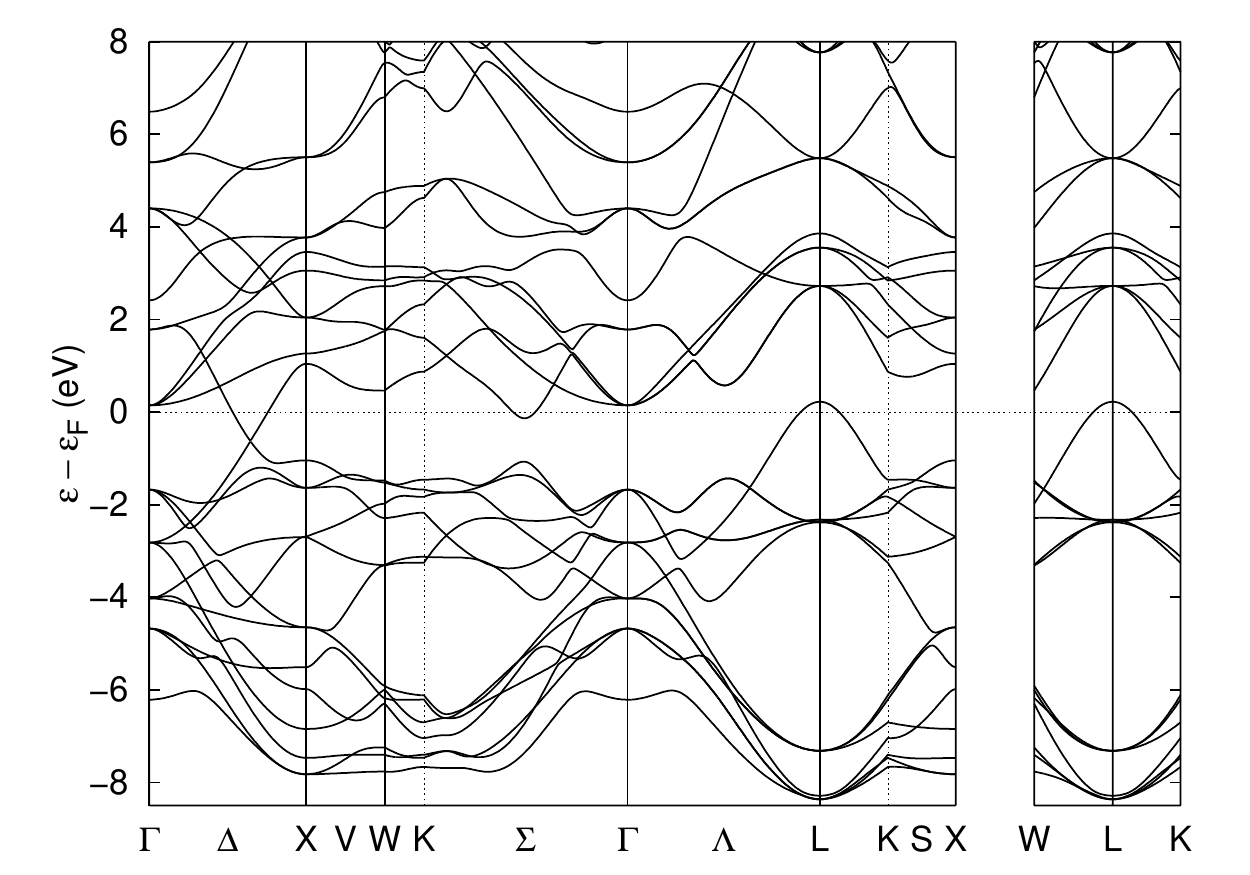} %NbO_bands}
  \caption{\small \label{fig:nboband} The electronic band structure of WN
    in the NbO structure at the PBE equilibrium (a = 4.131 \AA).}
\end{figure}

\begin{figure}
  \includegraphics[width=8.5cm]{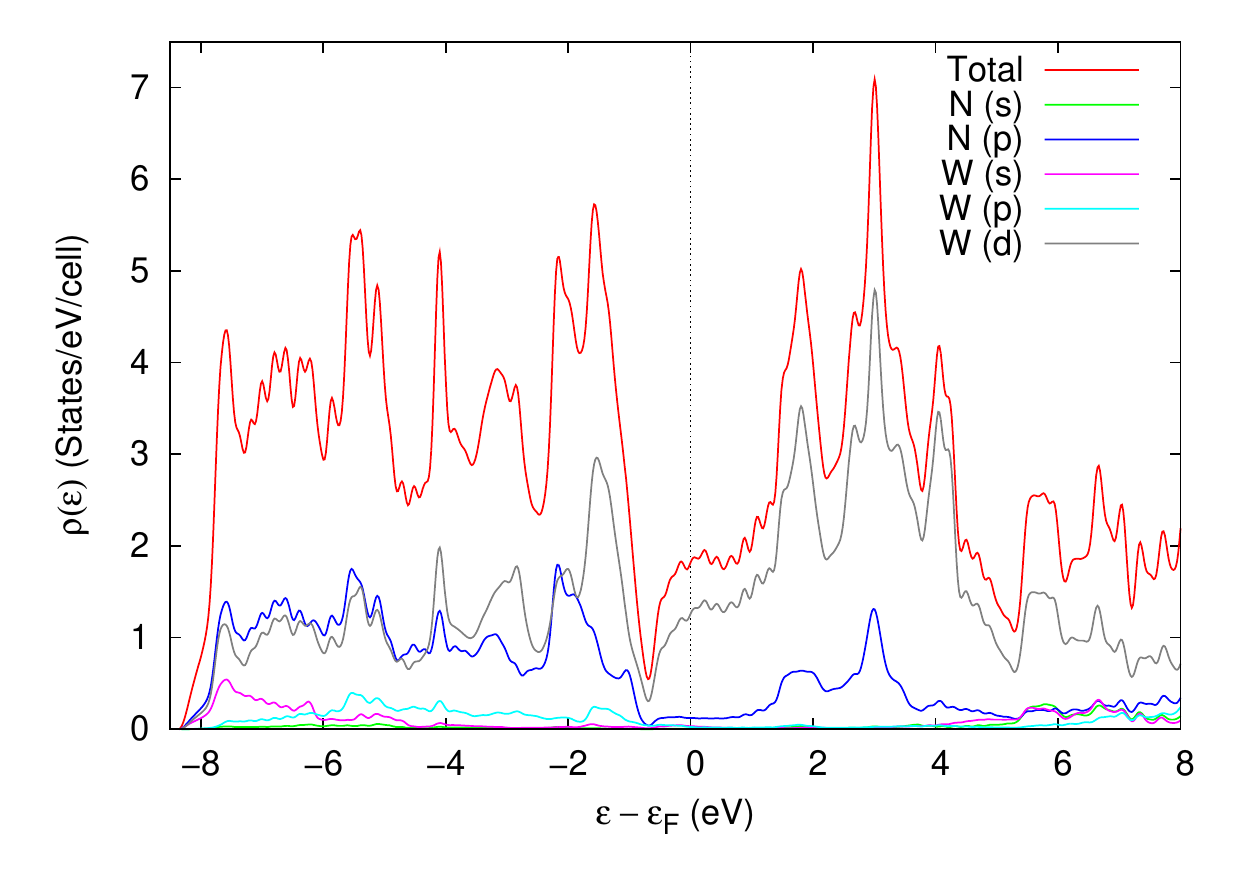} %NbO_dos}
  \caption{\small \label{fig:nbodos} (Color Online) The electronic density
    of states and angular-momentum decomposed density of states for
    WN in the NbO structure. These were computed via the tetrahedron method in
    VASP. The primary contributions are from the N-p and W-d
    states, with the W-d states dominating the conduction band.}
\end{figure}

The electronic band structure and density of states of the NbO
structure of WN are given in Figures \ref{fig:nboband} and
\ref{fig:nbodos}, respectively. The structure is metallic, with the
tungsten $d$ bands dominating the region near and above the Fermi
level.

The prediction of the hitherto unseen NbO ground state was
unexpected (albeit perhaps unsurprising in retrospect), and could
possibly be an artifact of either the PAW potentials or the PBE
exchange-correlation approximation. Accordingly, we compared the
energy of the NbO phase of WN to several of the usual suspects for
the WN composition. We used two techniques:
\begin{enumerate}
\item All electron calculations using the Elk FP-LAPW
  code\cite{dewhurst:elk} within the PBE GGA. Elk does not easily do
  structural relaxation, so we used the equilibrium structural
  parameters from the VASP PBE GGA calculations and the same k-point
  mesh. We used the supplied muffin-tin files for W and N, and set
  the energy cutoff using a value of RG$_{max}$ = 9.0.
\item LDA\cite{ceperley80:egas,perdew81:_sic} calculations using the
  LDA PAW potentials supplied with VASP. In this case we allowed
  the structures to fully relax. All other parameters, including
  the energy cutoff, we kept the same as in the PBE GGA
  calculations.
\end{enumerate}

\begin{table}
  \caption{\small \label{tab:equcomp} Energy differences for various
    structures of WN with 50-50 stoichiometry, computed using VASP
    with PBE PAW, the Elk FP-LAPW code, and VASP with LDA PAW. The
    Elk calculations used the relaxed structures from the PBE PAW
    runs, while the VASP LDA calculations were fully relaxed. In
    each case the energy of the NbO structure was set to zero, and
    we show energy differences between the structures in eV/atom.
    In the B8$_1$ (NiAs) calculation the N atom was placed on the As
    (Wyckoff position 2c) site.}
  \begin{tabular}{c|c|c|c}
    Structure & VASP PBE & Elk & VASP LDA \\
    \hline
    NbO & 0 & 0 & 0 \\
    B8$_1$ (NiAs) & 0.219 & 0.191 & 0.124 \\
    B$_h$ (WC) & 0.315 & 0.303 & 0.233 \\
    B1 (NaCl) & 0.590 & 0.591 & 0.603 \\
    B3 (ZnS) & 0.723 & 0.735 & 0.637
  \end{tabular}
\end{table}

Table~\ref{tab:equcomp} shows the results. The energy differences
between the VASP and Elk PBE calculations are all less than
0.02~eV/atom. The VASP LDA results show the same ordering of the
structures, although the energy differences are larger (as much as
0.1~eV/atom). We conclude that the NbO state is the lowest energy
structure of all the 50-50 stoichiometries structures we have
studied.

Since the NbO ground state WN can be considered as an NaCl structure
with ordered defects, it is logical to ask if any other pattern of
defects produce low energy structures. As mentioned at the beginning
of this section, the
literature\cite{wriedt89:WNphases,hagg30:WN,schonberg54:MoNWN,chiu93:WNfilms}
suggests that $\beta$WN is Nitrogen deficient, with composition
approximately WN$_{0.7}$. The simplest Nitrogen deficient structure
we can make is to take one atom out of an 8 atom supercell of the
NaCl structure, or, equivalently, add one Tungsten atom back into
the NbO structure. Experimentally, the prototype for this structure
is S$_3$U$_4$,\cite{zumbusch40:S3U4} and so we will refer to it by
that designation. As shown in
Figs.~\ref{fig:pbehull}-\ref{fig:vdwhull}, the enthalpy/atom of
W$_4$N$_3$ is lower in the S$_3$U$_4$ structure than the enthalpy of
WN in the NaCl structure, but it is still significantly larger than
the enthalpy of WN in the NbO structure. As a test, we also studied
Nitrogen-rich W$_3$N$_4$ in the S$_3$U$_4$ structure. We found that
its enthalpy was also lower than NaCl -- and even lower than
W$_4$N$_3$ in the S$_3$U$_4$ structure -- but still well above the
enthalpy of the NbO structure.

\begin{figure}
  \includegraphics[width=8cm]{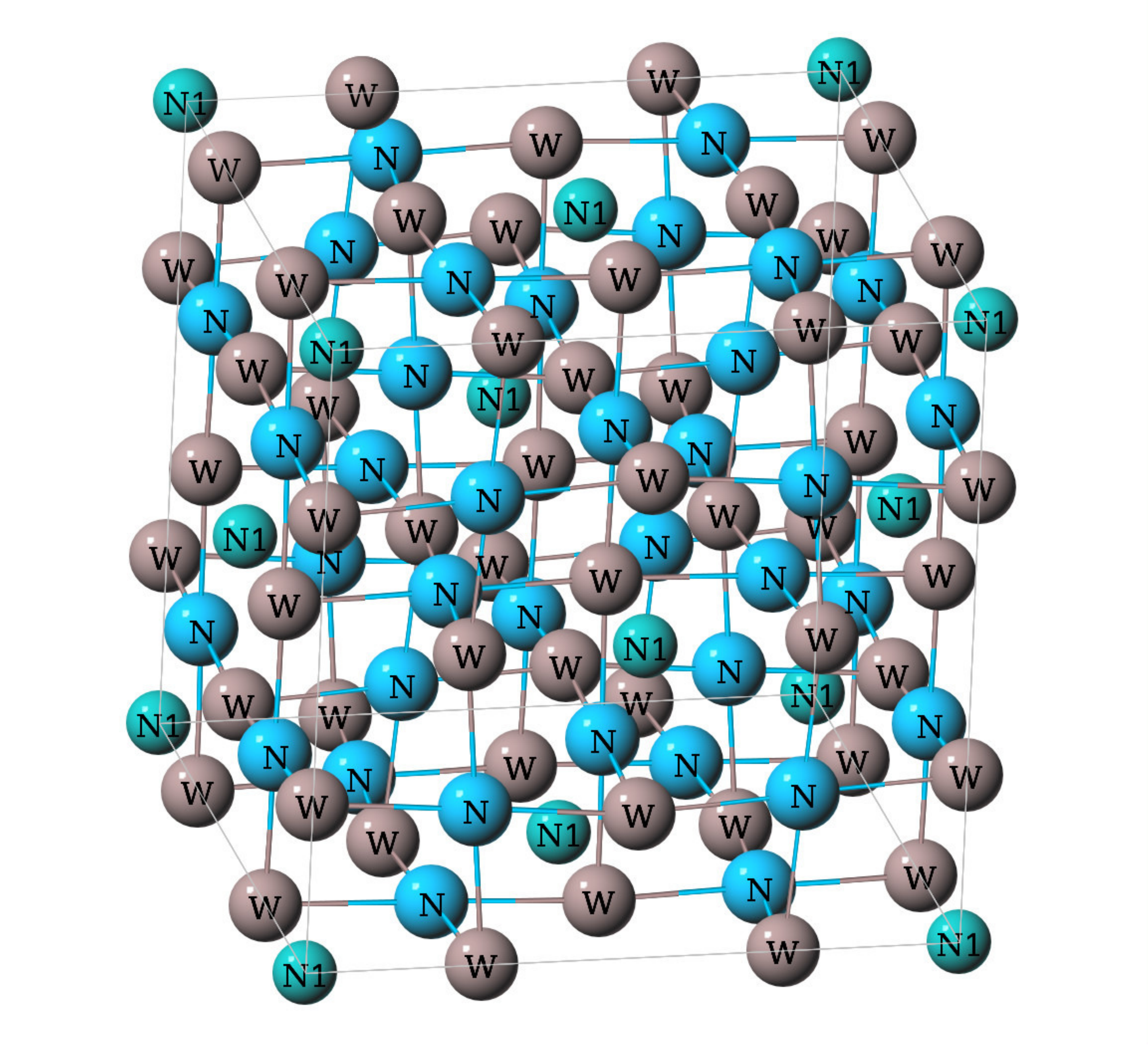} %{W6N7}
  \caption{\small \label{fig:n7w6} (Color Online) The N$_7$W$_6$ supercell.
    The atoms labeled N and W form the NbO structure. The atom
    labeled N1 was added at one of the vacancy sites, maintaining
    cubic symmetry. Note the substantial relaxation of the W atoms
    away from the N1 atom. Aside from NbO this is the lowest
    enthalpy structure of all those studied.}
\end{figure}

We continued along this path by constructing vacancy patterns in
NaCl supercells containing 8, 16, and 32 atoms. We considered a
variety of possible unit cells where a given site in the supercell
was ``on'' (occupied) or ``off'' (vacancy). We then fully relaxed
the unit cell, and computed the enthalpy according to
(\ref{equ:enthalpy}).  For an N atom cell there are $2^N - 1$
possible combinations. For the larger cells we have therefore currently
placed constraints on the search algorithm such that the composition is
WN$_x$ with $x \in [1/2,3/2]$. All of these structures can be
considered as candidates for H\"{a}gg's $\beta$ phase.

For the 8 atom supercell (a simple cubic supercell of NaCl) we
looked at all 255 combinations. We found 34 unique structures,
including NbO itself, CsCl, S$_3$U$_4$, ReO$_3$,\cite{chang78:reo3}
cubic perovskite (with formula unit NW$_4$ or WN$_4$), monatomic fcc
({\em Strukturbericht} symbol A1), and simple cubic (A$_h$), as well
as a variety of lower symmetry structures.

The 16 atom supercell is face-centered cubic, with all primitive
vectors doubled from the original NaCl cell. We looked at 26
structures with composition W$_{m+4}$N$_{n+4}$, where $m,n \ge
4$.

The 32 atom supercell of the NaCl structure is body-centered
cubic. In this case we only looked at structures closely related to
the NbO structure. Thus we examined W$_{13}$N$_{12}$, which has one
of the vacant W sites in the NbO structure occupied, and
W$_{12}$N$_{11}$, which has an extra N atom removed from the NbO
supercell. We looked at 18 structures of varying composition. Aside
from the near 50-50 stoichiometric structures, the most interesting
one has the composition N$_{12}$W$_6$. This structure will be
described in detail in Sec.~\ref{sec:WN2}.

The low enthalpy structures generated by this process are plotted as
red squares in Fig.~\ref{fig:hull}. All are closely related to NbO,
either by the addition or subtraction of an additional N or W atom
from the 16 or 32 atom supercells. For example, the lowest energy
structure after NbO plotted on the graph has composition
N$_7$W$_6$. We constructed this structure starting with the NbO
structure expanded onto the 16 atom supercell. We then placed an
additional N atom back onto one of the vacant sites, and allowed the
system to relax. The cell is shown in Fig.~\ref{fig:n7w6}.  There is
every reason to suspect that with increasing supercell size we can
find a large number of structures with enthalpy close to NbO, with
varying compositions.

Our calculations show that low-lying $\beta$ structures must have
vacancies on both the N and W sites. Since there are a large number
of these structures very close to the NbO state, entropic effects
must be large. The observed ``sodium chloride'' structure, with
lattice constant 4.12-4.14\AA,\cite{wriedt89:WNphases,khitrova59:WN}
is most likely formed with approximately 25\% vacancies on each
site. These vacancies will no doubt be ordered locally, but when
averaged over the entire crystal will look like the NaCl structure.
The lattice constant of this phase will be closer to NbO (4.131\AA)
than pure NaCl (4.366\AA). The experimentally observed lattice
constant of 4.12-4.14\AA ~is consistent with this conjecture.

\section{The hexagonal $\delta_H$ structures \label{sec:deltanw}}

Khitrova and Pinsker\cite{khitrova62:WN} detailed six hexagonal
``$\delta_H$'' and rhombohedral ``$\delta_R$'' phases in the W-N
system, with compositions ranging from NW$_2$ ($\delta_H^{II}$) to
N$_2$W ($\delta_H^{III}$). Four of these structures were found to
have vacancies on one of the tungsten sites. Since we were primarily
interested in the nitrogen-rich structures close to the convex hull
we did not do exhaustive searches for ordered vacancies in these
structures. We did, however, find that the $\delta_H^I$ structure,
taken without defects and thus having composition N$_2$W$_3$, had
the lowest energy of all the $\delta$ structures.
Fig.~\ref{fig:hull} shows that this structure is within 0.1~eV/atom
of the tie-line within the PBE, and the structure is barely above
the tie-line within the LDA (Fig.~\ref{fig:ldahull}). It is quite
likely that searches for ordered defect structures in supercells of
$\delta_H^I$ will show lower energies.

\section{Predicted WN$_2$ structures \label{sec:WN2}}

In both the LDA and PBE the lowest lying states with composition
WN$_2$ are the hexagonal structures found by Wang {\em et
  al.}\cite{wang09:wn2}. Our structural parameters agree with
theirs, as shown in the early columns of Table~\ref{tab:hexN2W}. We
also found the elastic constants of both structures, including the
average of the Voigt\cite{voigt28:GV} and Reuss\cite{reuss29:GR}
bounds on the isotropic bulk and shear moduli. These also agree with
Ref.~\onlinecite{wang09:wn2}, confirming that these structures are
candidates for superhard WN.

\begin{table}
  \caption{\small Equilibrium structural parameters for the hexagonal
    N$_2$W structures found by Wang {\em et al.}\cite{wang09:wn2}
    The low symmetry hP3 structure is in space group
    $P\overline{6}m2$-$D_{3h}^1$ (\#187), with nitrogen atoms at the
    (2g) Wyckoff positions $(00z)$ and the tungsten atom on the (1d)
    site $(\frac13 \frac23 \frac12)$. In the high symmetry hP6
    structure ($P6_3/mmc$-$D_{6h}^4$, \#194) the nitrogen atoms are
    on (4e) sites $(00z)$, and the tungsten atoms are on (2d) sites
    $(\frac13 \frac23 \frac34)$. For the LDA and PBE functionals we
    show our results (O), and the results of Wang {\em et al.}
    (W). R$_{N-N}$ and R$_{N-W}$ are the length of the respective
    bonds. The elastic constants are in GPa. The isotropic bulk
    ($B$) and shear ($G$) moduli are averages of the
    Voigt\cite{voigt28:GV} and Reuss\cite{reuss29:GR}
    bounds. \label{tab:hexN2W}}
  \begin{tabular}{c|cc|cc|c}
    \hline\hline
    Functional & \multicolumn{2}{c|}{LDA} & \multicolumn{2}{c|}{PBE} &
    vdW-DF2 \\
    \hline\hline
    \multicolumn{6}{c}{Low symmetry $P\overline{6}m2$, hP3} \\
    \hline
    & O & W & O & W & O \\
    \hline
    a (\AA) & 2.887 & 2.887 & 2.928 & 2.933 & 3.000 \\
    c (\AA) & 3.877 & 3.877 & 3.916 & 3.918 & 3.974 \\
    z & 0.1804 & 0.1804 & 0.1813 & 0.1814 & 0.1824 \\
    R$_{N-N}$ (\AA) & 1.399 & 1.399 & 1.421 & 1.421 & 1.450 \\
    R$_{N-w}$ (\AA) & 2.077 & 2.077 & 2.104 & 2.104 & 2.143 \\
    \hline
    $C_{11}$ & 638 & 654 & 573 & 588 & 493 \\
    $C_{33}$ & 1056 & 1082 & 954 & 973 & 827 \\
    $C_{44}$ & 241 & 260 & 222 & 232 & 191 \\
    $C_{12}$ & 207 & 213 & 183 & 191 & 159 \\
    $C_{13}$ & 230 & 248 & 193 & 206 & 157 \\
    $B$     & 396 & 412 & 351 & 255 & 297 \\
    $G$     & 246 & 255 & 226 & 231 & 195 \\
    \hline\hline
    \multicolumn{6}{c}{High symmetry $P6_3/mmc$, hP6} \\
    \hline
    & O & W & O & W & O \\
    \hline
    a (\AA) & 2.893 & 2.893 & 2.934 & 2.939 & 3.007 \\
    c (\AA) & 7.714 & 7.714 & 7.790 & 7.796 & 7.891 \\
    z & 0.0898 & 0.0898 & 0.0902 & 0.0902 & 0.0907 \\
    R$_{N-N}$ (\AA) & 1.392 & 1.392 & 1.406 & 1.406 & 1.431 \\
    R$_{N-w}$ (\AA) & 2.078 & 2.078 & 2.105 & 2.105 & 2.144 \\
    \hline
    $C_{11}$ &  633 &  642 & 568 & 579 & 486 \\
    $C_{33}$ & 1051 & 1078 & 952 & 973 & 827 \\ 
    $C_{44}$ &  249 &  262 & 228 & 233 & 197 \\
    $C_{12}$ &  213 &  217 & 187 & 195 & 161 \\
    $C_{13}$ &  237 &  252 & 199 & 211 & 197 \\
    $B$     &  399 &  411 & 353 & 364 & 298 \\
    $G$     &  245 &  252 & 225 & 228 & 195 \\
    \hline
  \end{tabular}
\end{table}

Both of these structures can be described as N$_2$ dimers separated
by Tungsten planes. We might therefore expect the addition of van
der Waals forces between the dimers to be significant. Structurally,
as shown in the last column of Table~\ref{tab:hexN2W}, there is
little difference between the vdW-DF2 calculations and the LDA or
PBE. The hexagonal unit cell is expanded slightly in both
structures, as is the N$_2$ dimer separation, but the changes are
not substantial. The energetic difference, is, however, quite
large. In the LDA (Fig.~\ref{fig:ldahull}) and PBE
(Fig.~\ref{fig:pbehull}) these WN$_2$ structures have
enthalpies/atom comparable to the NbO structure of WN.

\begin{figure}
  \includegraphics[width=7cm]{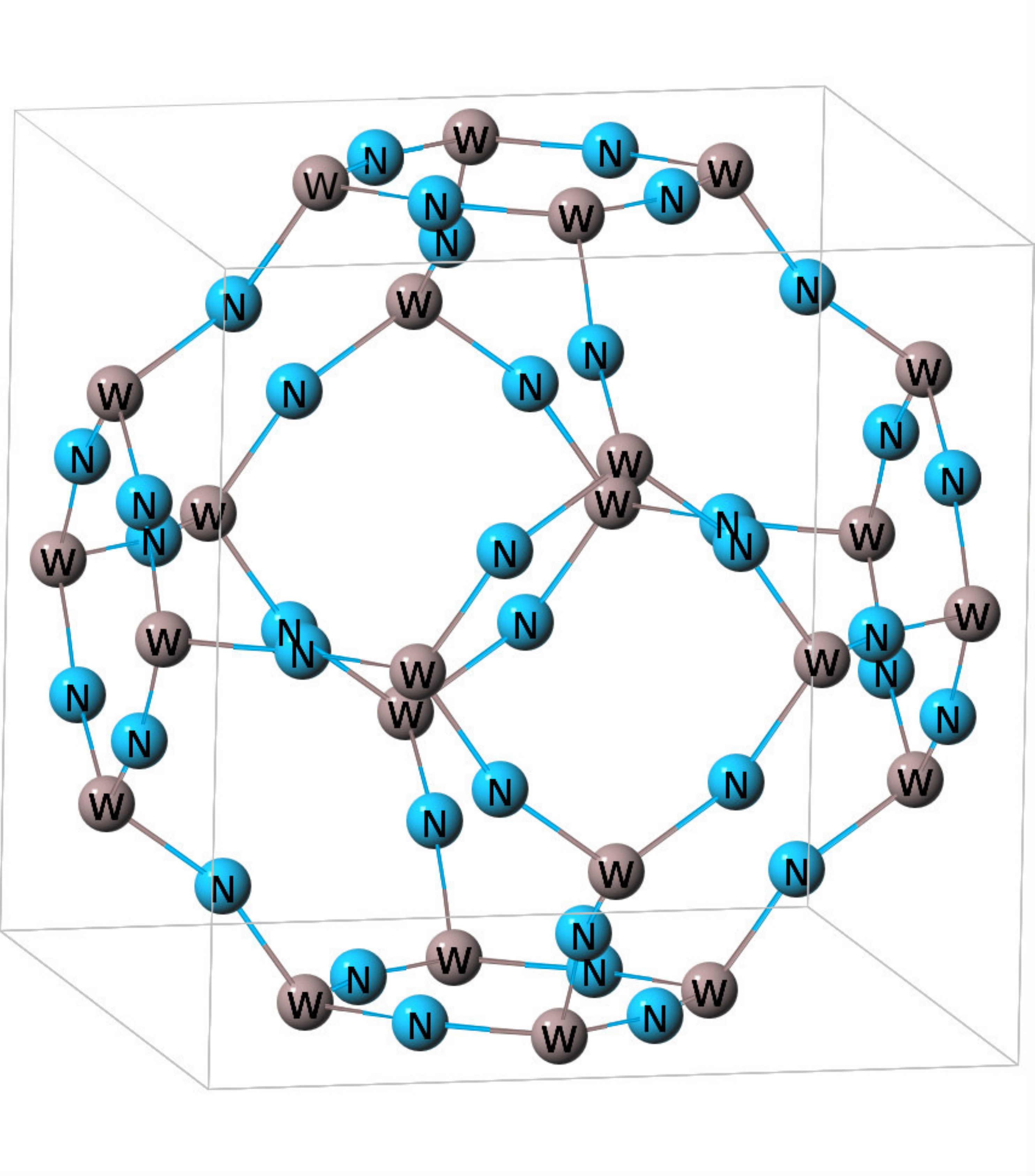} %cI36}
  \caption{\small (Color online) The cI36 structure predicted for
    N$_2$W. The basic structure looks the same using LDA, PBE, and
    vdW-DF2 functionals. The exact dimensions are described in
    Table~\ref{tab:cI36}. Note that this is a body-centered cubic
    lattice, so the center of the cube and the cube corners are
    equivalent sites.\label{fig:cI36}}
\end{figure}

\begin{table}
  \caption{\small The cI36 structure of WN$_2$, constructed from a 32 atom
    body-centered cubic supercell of the NaCl structure. The space
    group is $Im\overline{3}m$-$O_h^9$ (\#229), the nitrogen atoms
    occupy the (24h) Wyckoff positions $(0yy)$, and the tungsten
    atoms occupy the (12d) sites $(0 \frac{1}{4} \frac{1}{2})$. The
    first N-W-N angle is for atoms lying on the surface of the cube
    cell. The second case has one of the N atoms in the interior of
    the cube.\label{tab:cI36}}
  \begin{tabular}{c|ccc}
\hline\hline
  Functional  & LDA & PBE & vdW \\
  \hline
  a (\AA) & 10.273 & 10.383 & 10.490 \\
  y & 0.1454 & 0.1455 & 0.1455 \\
  W-N bond (\AA) & 1.840 & 1.860 & 1.879 \\
\hline
\multicolumn{4}{c}{Angles} \\
\hline
W-N-W & 108.7$^\circ$ & 108.6$^\circ$ & 108.5 $^\circ$ \\
N-W-N (face) & 108.7$^\circ$ & 108.6$^\circ$ & 108.5$^\circ$ \\
N-W-N (interior) & 109.9$^\circ$ & 109.9$^\circ$ & 110.0$^\circ$ \\
\hline
  \end{tabular}
\end{table}

Adding van der Waals forces changes the shape and composition of the
ground-state convex hull. The ground state structure of the system
is now the body-centered cubic structure shown in
Fig.~\ref{fig:cI36}.  Although it is not obvious, this structure was
constructed from a 32-atom body-centered cubic supercell of
NaCl. This results in a structure with space group
$Im\overline{3}m$-$O_h^9$, \#299. Nitrogen atoms are at the (2a),
(6b), and (24h) Wyckoff positions of the BCC lattice, while tungsten
atoms are on the (8c), (12d), and (12e) sites. We remove the atoms
on the (2a), (6b), (8c), and (12e) sites, leaving 12 N atoms at the
$(0,y,y) (24h)$ Wyckoff positions, and 6 W atoms on Wyckoff sites
$(0 1/4 1/2) (6d)$. The nitrogen atoms are then allowed to relax
away from the $y = 1/4$ value of the original supercell. We have
relaxed this structure, which we will designate by its Pearson
symbol, cI36, using all three functionals. The results are shown in
Table~\ref{tab:cI36}.

Within the LDA the cI36 structure is not particularly noteworthy.
Its enthalpy is above the tie-line, well above that of our other
WN$_2$ structures, including Mo$_2$N, another structure which can be
formed by removing atoms from the NaCl structure.  With the PBE
functional the cI36 structure is less than 0.1~eV/atom above the
ground state hull, below all of our test structures save for the two
from Wang~{\em et al.}

Finally, when we use the vdW-DF2 functional the cI36 structure
becomes the lowest energy structure. This is not necessarily because
the van der Waals interaction promotes this structure. Rather, as is
seen in Fig.~\ref{fig:vdwhull}, the energy of the hexagonal WN$_2$
structures moves up compared to the $\alpha$N$_2$ and bcc W
endpoints, even though the lattice constants, atomic positions, and
bond lengths are little changed. Since there is no experimental
evidence for any of the WN$_2$ structures, and all three functionals
are approximations with well-known limitations, we cannot determine
which picture is correct.

\begin{figure}
  \includegraphics[width=8.5cm]{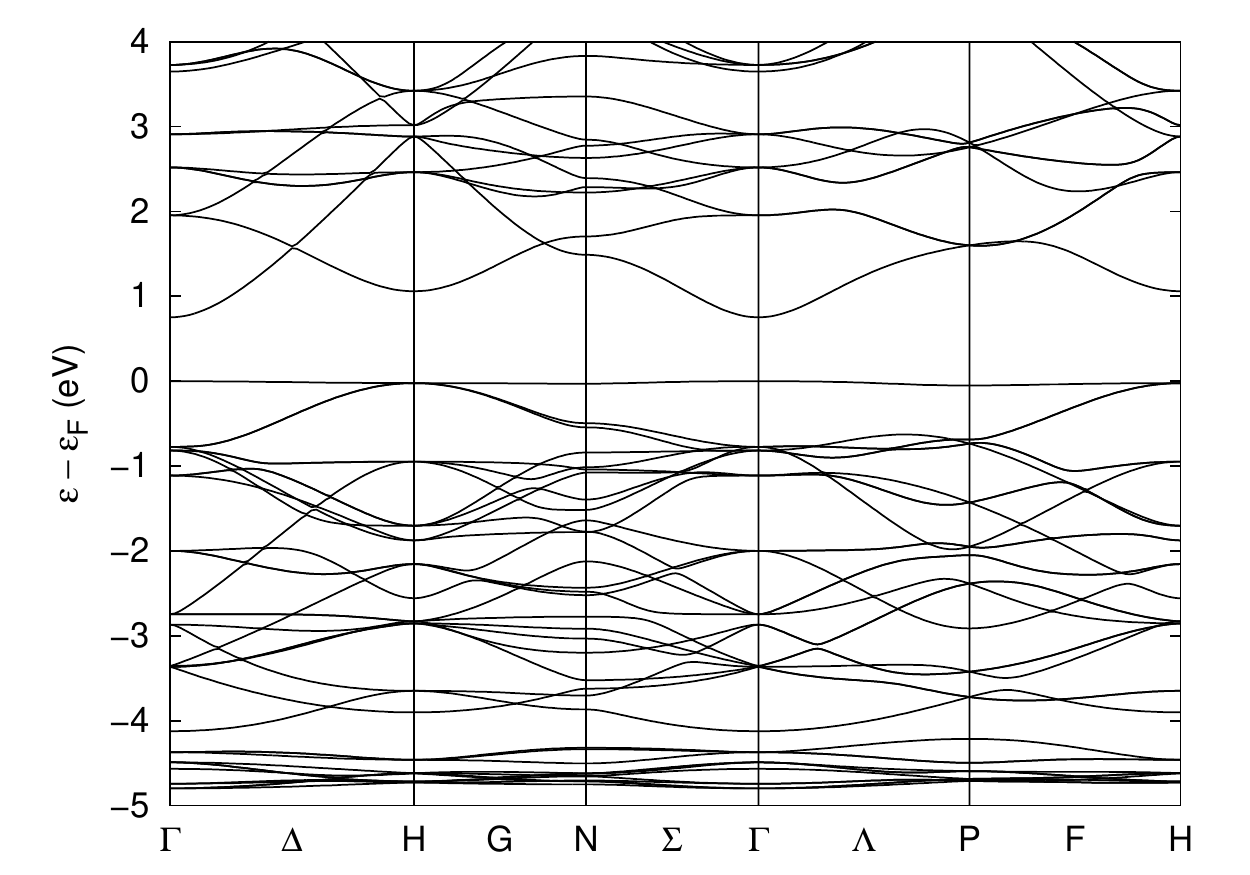} %cI36vdW_bands}  
  \caption{\small The electronic band structure for the body-centered cubic
    cI36 structure of WN$_2$ described in the text. Calculations
    were done using the vdW-DF2 functional, including van der Waals
    forces, within VASP. The top valence band has a very small
    dispersion, dropping to -0.05~eV below the Fermi level at P. The
    gap is 0.75~eV, and is direct at $\Gamma$. The LDA and PBE
    calculations produce similar band
    structures.\label{fig:ci36band}}
\end{figure}

The cI36 structure is an insulator. The electronic band structure,
obtained using the vdW-DF2 functional, is shown in
Fig.~\ref{fig:ci36band}. The phase is also elastically stable, as is
shown in Table~\ref{tab:ci36cij}. We computed the elastic constants
using the finite-strain method,\cite{mehl90:_struc} fixing the
primitive cell in its strained position while allowing the atoms to
relax within the cell, as allowed by symmetry. Whereas cubic WN, in
the NbO structure, is quite hard, the shear moduli of the cI36
structure is predicted to be very small by all functionals.

\begin{table}
  \caption{\small Elastic constants for the cI36 structure of WN$_2$,
    computed using the LDA, PBE, and vdW-DF2 functionals. Elastic
    constants were computed by the finite strain
    method,\cite{mehl90:_struc} allowing the atoms to relax at each
    strain while fixing the unit cell. Equilibrium structural
    parameters are taken from Table~\ref{tab:cI36}. $B$ is the
    equilibrium bulk modulus, and all elastic constants are in
    GPa.\label{tab:ci36cij} The shear modulus $G$ is an average of
    the Hashin-Shtrikman bounds.\cite{hashin62:var,hashin62:poly}}
  \begin{tabular}{c|ccc|ccc}
    \hline\hline
    Functional  & $C_{11}$ & $C_{12}$ & $C_{44}$ & $B$ & $C_{11}$ -
    $C_{12}$ & $G$ \\
    \hline
    LDA     & 119 & 77 & 13 & 91 & 42 & 15 \\
    PBE     & 112 & 72 & 13 & 85 & 41 & 16 \\
    vdW-DF2 & 104 & 65 & 12 & 78 & 39 & 14
  \end{tabular}
\end{table}

Further work on possible WN$_2$ phases was done by Song and
Wang,\cite{song10:wn} who studied WN$_2$ in the CoSb$_2$
structure.\cite{siegrist86:CoSb2} As shown in
Figs.~\ref{fig:pbehull}-\ref{fig:vdwhull}, this structure is
0.1-0.2~eV above the hexagonal WN$_2$ structures for all three
functionals. It is therefore never a candidate for the W-N ground
state. We also found two other structures with energies close to the
CoSb$_2$ structure: Mo$_2$N,\cite{evans57:mo2n} which is another
structure constructed by removing atoms from a supercell of NaCl,
and AuSn$_2$.\cite{rodewald06:AuSn2} Neither structure is a
candidate for the ground state of WN$_2$. Parameters of the minimum
energy structures for all of these phases are given in the
supplementary material.

\section{WN$_3$ Structures\label{sec:WN3}}

Song and Wang\cite{song10:wn} considered WN$_3$ in the P$_3$Tc
structure,\cite{ruhl82:P3Tc} and found that it has a negative
enthalpy relative to $\alpha$N$_2$ and bcc-W. They showed that the
elastic constants met the Born criteria for long-wavelength
stability,\cite{born40:stability} and so claim that it is a possible
state of WN$_3$.  As seen in Fig.~\ref{fig:ldahull}, within the LDA
the P$_3$Tc does have negative enthalpy compared to the end
points. However, it is over 0.2~eV/atom above the tie-line using the
LDA, PBE, and vdW-DF2 functionals. This means that the PTc$_3$
structure cannot be an equilibrium state in the W-N system, although
it is possible it may exist as a metastable state.

We also looked at a competing structure, Molybdite
(MoO$_3$).\cite{andersson50:MoO3} This structure has the same space
group ($Pnma$) as P$_3$Tc and the same occupied Wyckoff positions.
Yet when starting from different initial conditions it relaxes to a much
different unit cell, even though the energy is comparable to the
PTc$_3$ structure. The minimum-energy structures for both P$_3$Tc
and MoO$_3$ are given in the supplementary material.

\begin{figure}
  \includegraphics[width=8.5cm]{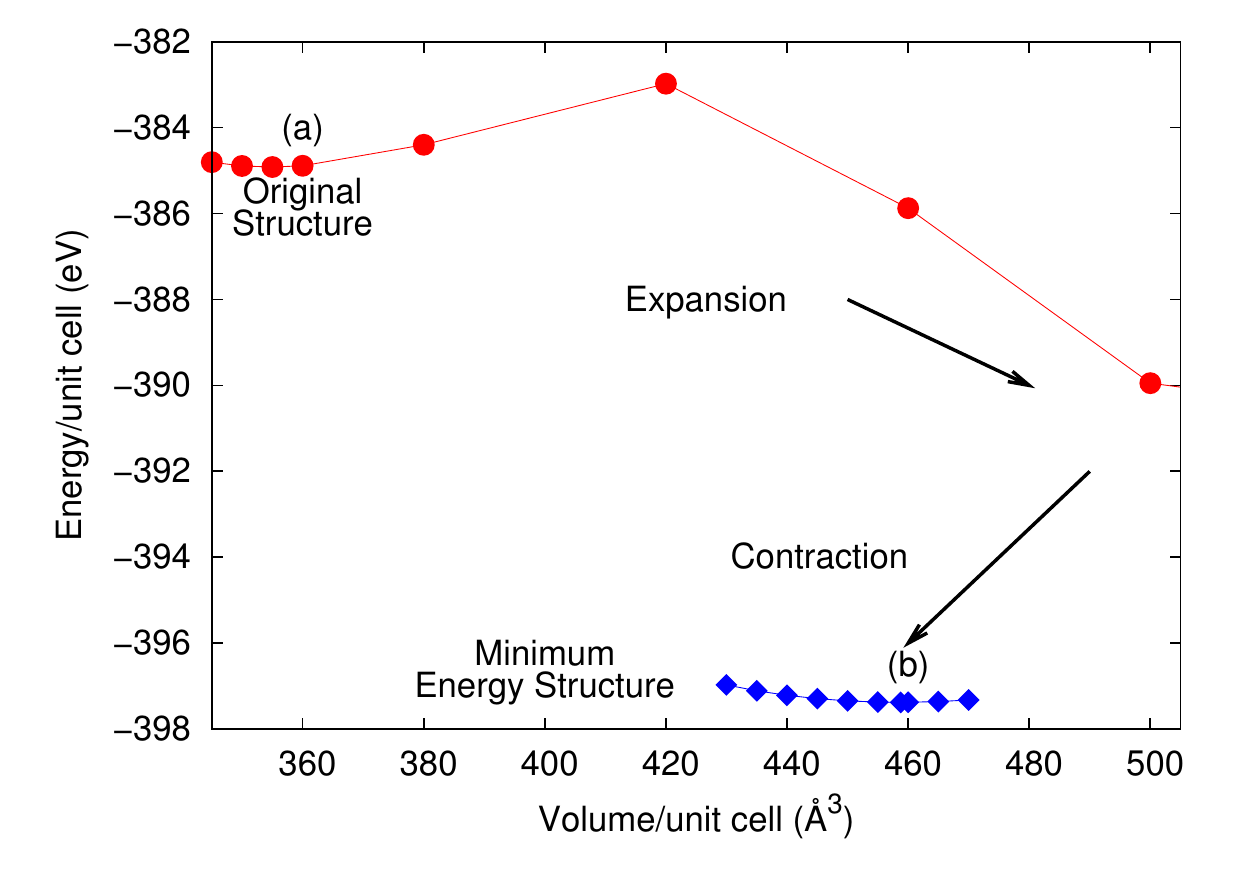} %ReP4}
  \caption{\small (Color online) Energy versus volume calculations for
    WN$_4$ in the ReP$_4$ structure, using VASP and the LDA
    exchange-correlation functional. The original calculation in the
    upper left-hand corner found a structure close to that reported
    previously.\cite{aydina12:XN4} When the structure was expanded
    past a volume of 420~\AA$^3$, the energy started to drop
    rapidly. Reversing the expansion caused another shift in the
    structure, where the energy finally reached the values shown
    near the bottom of the graph. The labels (a) and (b) correspond
    to the structure pictures in Fig.~\ref{fig:ReP4struc}. All the
    calculations use the same space group, $Pbca$-$D_{2h}^{15}
    (\#61)$, and the N and W atoms remained on (8c) Wyckoff
    positions.\label{fig:ReP4lda}}
\end{figure}

\begin{figure*}
  \includegraphics[width=8cm]{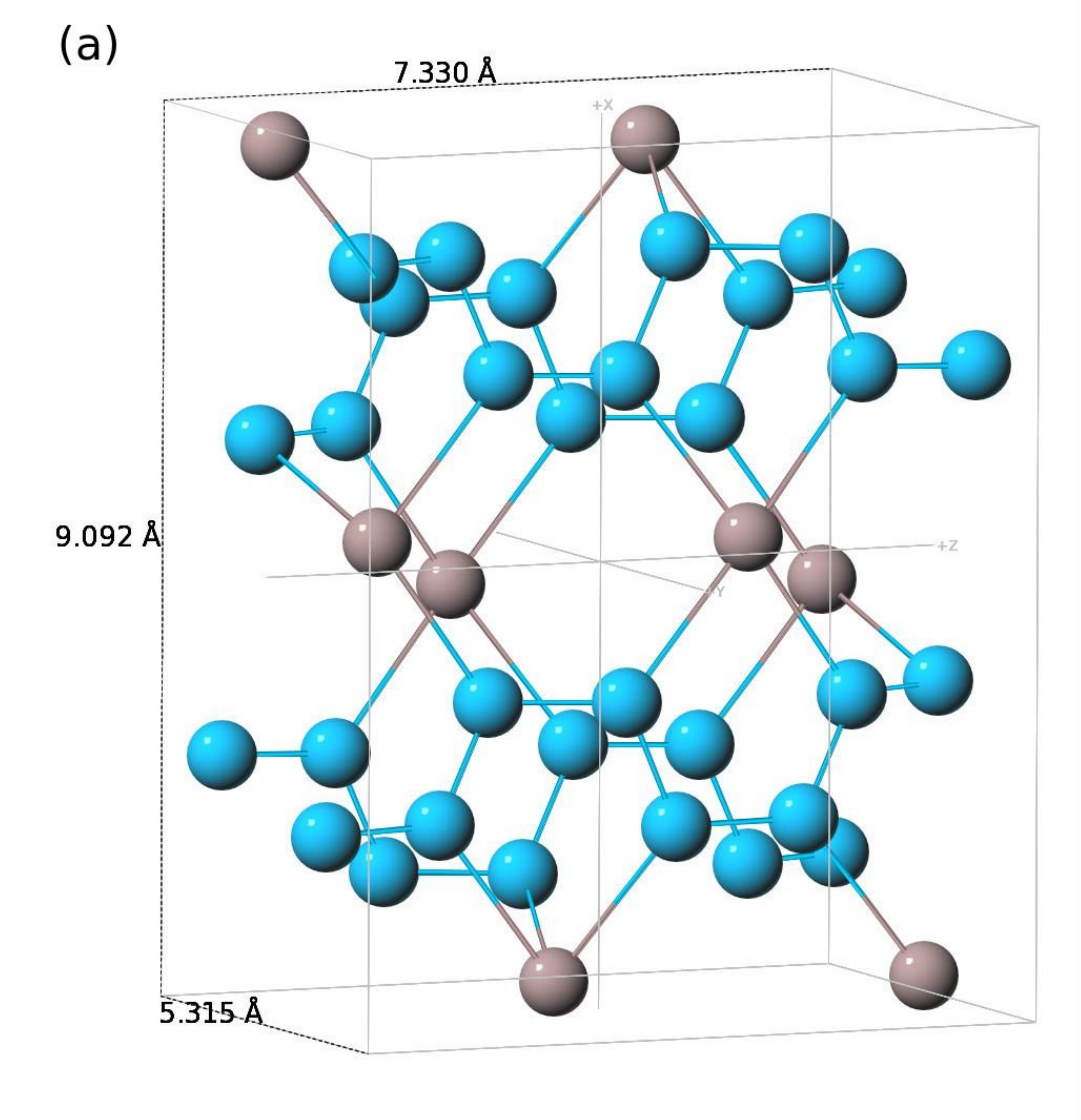} %ReP4_LDA_H}
  \includegraphics[width=8cm]{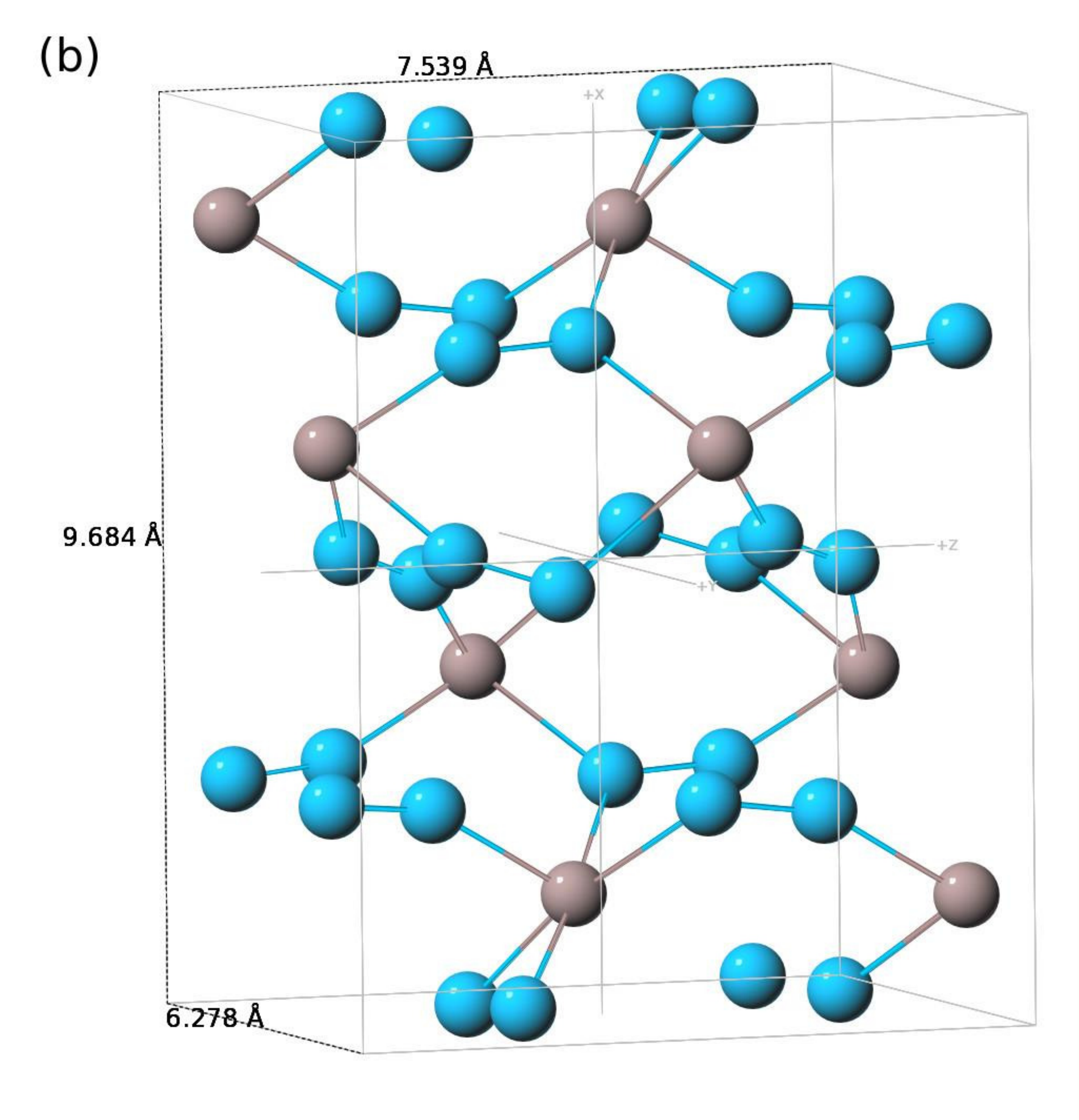} %ReP4_LDA_L}
  \includegraphics[width=8cm]{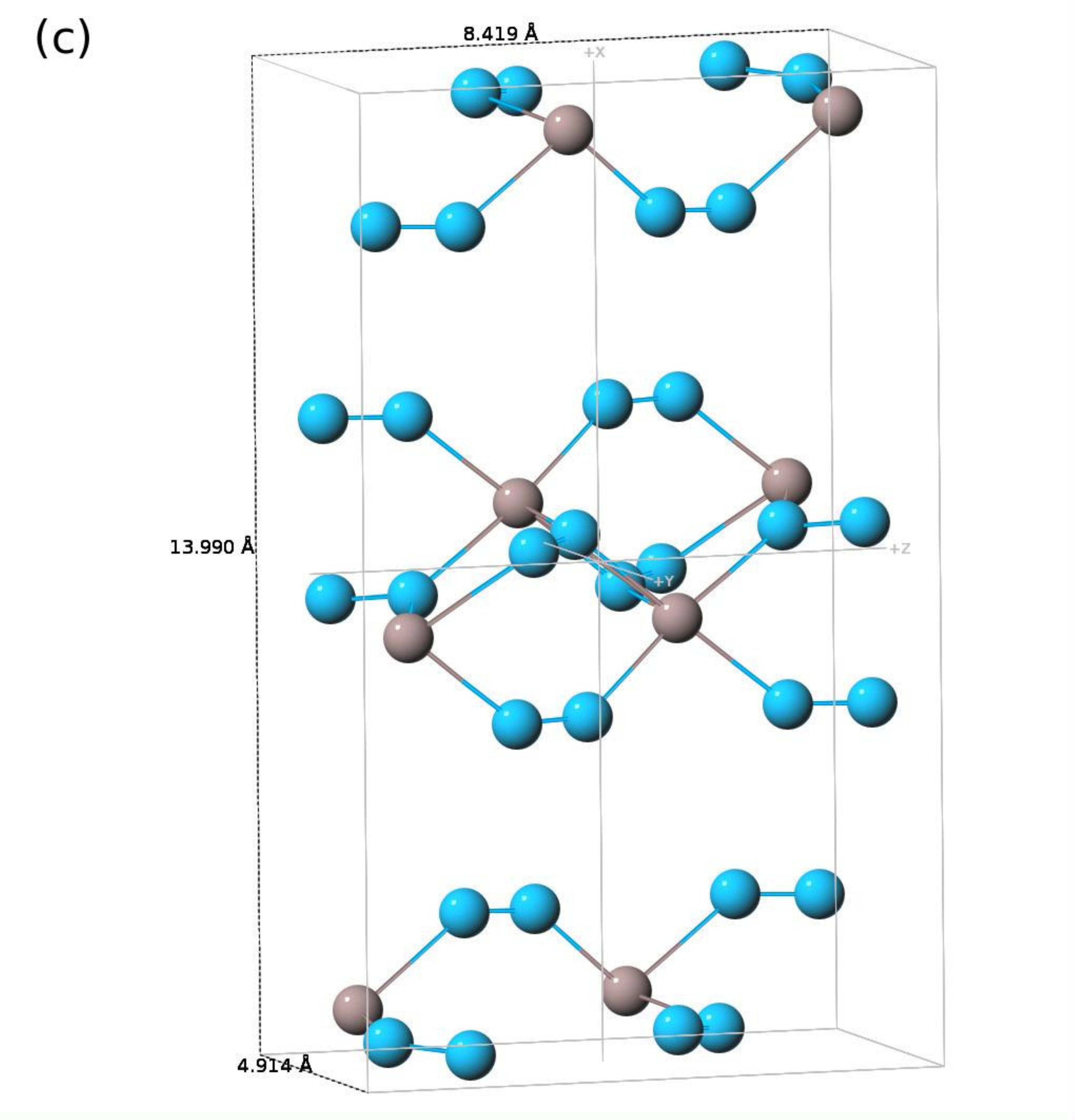} %ReP4_PBE}
  \includegraphics[width=8cm]{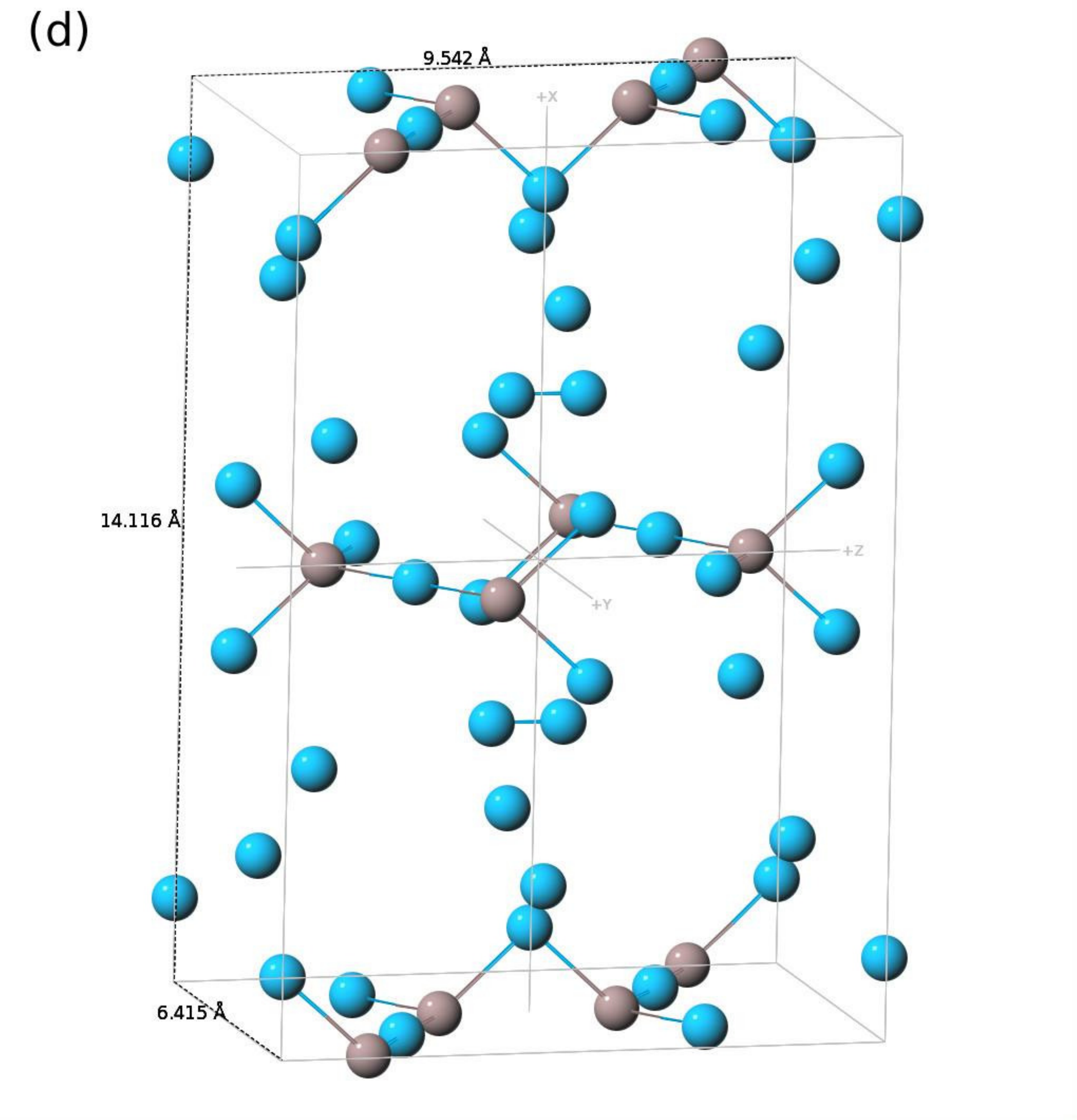} %ReP4_vdWDF2}
  \caption{\small Possible metastable primitive cells of N$_4$W in the
    ReP$_4$ structure. The distances are the values of the
    orthorhombic lattice constants $(a,b,c)$ which locally minimize
    the total energy. The LDA functional was used for (a) and (b),
    which correspond the the structures referred to in
    Fig.~\ref{fig:ReP4lda}. The minimum energy structure for the PBE
    functional is shown in (c), and the minimum energy structure for
    the vdW-DF2 functional is shown in (d).  Minimum energy lattice
    and Wyckoff position parameters for all structures are given in
    the supplementary material.
    \label{fig:ReP4struc}}
\end{figure*}

\section{WN$_4$ Structure\label{sec:WN4}}

The WN$_4$ structure proposed by Aydin {\em et
  al.}\cite{aydina12:XN4} is the most interesting candidate
structure we studied. Using ReP$_4$ as its
prototype,\cite{jeitschko79:ReP4} within the LDA they predicted it
to have relatively large bulk and shear moduli (338 and 198 GPa,
respectively). In addition its hardness was estimated to be near
that of some of the superhard borocarbides and carbonitrides.

We investigated this structure in some detail. We relaxed the
structure provided by the authors of Ref.~\onlinecite{aydina12:XN4}, using 
the LDA, PBE, and vdW-DF2 structures. Within the LDA we quickly
found an equilibrium position close to the published
result. When we next computed the energy-versus volume curve, in
order to determine the bulk modulus, we inadvertently expanded the
structure past the unit cell volume of 420~\AA$^3$. The structure
quickly dropped in energy, as shown in
Fig.~\ref{fig:ReP4lda}. Compressing the lattice then caused a
transformation to yet another configuration, labeled (b) in
Fig.~\ref{fig:ReP4lda}.  Note that at no time did the structure
leave the $Pbca$ space group of ReP$_4$, and the atoms remained on
(8c) Wyckoff sites. Most likely the finite jumps in volume we
applied allowed for large relaxations in the atomic positions,
causing a jump from one local minimum to another. It is possible
that further manipulation of the structure will result in even lower
energies.  We compare the original structure (a) to the (current)
minimum energy LDA structure (b) in the first part of
Fig.~\ref{fig:ReP4struc}.

We also relaxed the original structure of Aydin using the PBE and
vdW-DF2 functionals. Both quickly relaxed away from the original
structure and into the results shown in Fig.~\ref{fig:ReP4struc}.  Note
that in the vdW-DF2 structure, the system has separated into layers
of WN$_2$ and N$_2$ molecules. It is unlikely that such a system
could be very hard.

\begin{table}
  \caption{\small Lattice constants and elastic constants for the WN$_4$
    structures shown in Fig.~\ref{fig:ReP4struc}. Elastic constants
    were computed within VASP using finite distortions of the unit
    cell, and include the internal relaxation of the atoms to the
    strain. Structure labels refer to the structures shown in
    Fig.~\ref{fig:ReP4struc}. Lattice constants are in
    {\AA}ngstr\"{o}ms, while elastic constants are in GPa.
    \label{tab:ReP4cij}}
  \begin{tabular}{ccccc}
    \hline\hline
    Structure & (a) & (b) & (c) & (d) \\
    Functional & LDA & LDA & PBE & vdW-DF2 \\
    \hline
    $a$ & 9.093 & 9.662 & 13.273 & 14.116 \\
    $b$ & 5.314 & 6.279 &  4.927 &  6.415 \\
    $c$ & 7.330 & 7.532 &  8.402 &  9.542 \\
    \hline
    $C_{11}$ & 742.1 &  86.8 &  12.1 &  14.6 \\
    $C_{22}$ & 159.2 &  96.6 & 138.9 & 131.3 \\
    $C_{33}$ & 660.1 & 232.6 & 279.7 & 132.8 \\
    $C_{44}$ &  23.7 & -11.9 &  55.5 &  20.5 \\
    $C_{55}$ & 383.0 &   7.5 &  -3.2 &  -4.7 \\
    $C_{66}$ &  74.4 &  87.8 &   2.8 &   1.0 \\
    $C_{12}$ & 102.5 &  42.3 &   4.6 &   5.5 \\
    $C_{13}$ & 314.6 &  58.8 &   8.0 &   2.9 \\
    $C_{23}$ &  47.3 &  82.4 &  69.4 &  92.0
  \end{tabular}
\end{table}

The minimum energy lattice constants and elastic constants for the
structures shown in Fig.~\ref{fig:ReP4struc} are given in
Table~\ref{tab:ReP4cij}. In the interest of saving space, we list
the atomic positions in the supplementary material. While the
structure labeled (a) has rather large elastic constants and so
might be considered a hard material, the other structures do
not. Since the (b) structure is lower in energy than the (a)
structure within the LDA, we conclude that the ReP$_4$ structure of
N$_4$W is not a hard material, assuming it can be made. We also note
that we cannot guarantee that any of the structures studied here is
the true minimum energy structure of ReP$_4$. There are too many
possible local minima to search through at this time.

The negative elastic constants for $C_{44}$ in the LDA (b) structure
and $C_{55}$ in the PBE and vdW-DF2 structures violate the Born
criteria for elastic stability. We are performing further
calculations on these systems in the hope of finding still lower
energy structures. Since we have already shown that the ReP$_4$
state is not a viable candidate for either a hard or a stable phase
in the WN system, we will defer discussion of these results.

\section{Discussion \label{sec:discuss}}

Modern first-principles electronic structure methods make it
relatively simple to determine the minimum energy configuration,
band structure, and total energy for any reasonably sized
structure. Once that is known, it is straightforward (although
frequently tedious) to determine if the structure is stable against
strains and vibrations, and thus a candidate for a stable or
metastable structure.

It is considerably harder to determine where a structure fits
energetically compared to other structures of similar
composition. In particular, we cannot determine if a given structure
is stable with respect to phase separation or a structural phase
transition until we determine the shape of the convex hull on the
enthalpy versus composition diagram. High-throughput methods such as
{\small AFLOW} make it simple to determine this hull. This approach can
expediently flag low-energy structures which might be candidates for
metastable states, as well as states which might be found via a
pressure or temperature driven phase transition.

This paper presents a high-throughput study of the tungsten-nitrogen
system.  Although there have been many first-principles calculations
for this
system,\cite{suetin08:wc-wn,wang09:wn2,song10:wn,aydina12:XN4,isaev07:tmcn,kroll05:TaN_WN}
there has never before been a study of possible structures with
multiple compositions. The high-throughput calculations were done
using {\small AFLOW}, which we enhanced by adding known nitride, oxide, and
boride structures which had not previously been included in the
database. These calculations confirmed the experimental observations
that tungsten nitride forms a cubic system based on the rock salt
structure with vacancies
($\beta$WN).\cite{hagg30:WN,khitrova59:WN,khitrova62:WN,khitrova62:WN2}

In fact, we found that the NbO structure (Fig.~\ref{fig:NbO}) is the
best candidate for the ground state structure, while other states
with varying patterns of vacancies produce structures nearly
degenerate with NbO-WN. Most attempts to form WN compounds used high
temperatures. It is thus likely that the experimental structure of
WN will be the NaCl structure with approximately 25\% random
vacancies on both the tungsten and nitrogen sites, with some local
ordering of the vacancies. This is not quite in agreement with
previous experiments, where it appears that the $\beta WN$ has
vacancies only on the nitrogen
site.\cite{hagg30:WN,khitrova59:WN,khitrova62:WN,khitrova62:WN2} The
experimental literature on the WN system is sparse, and more data is
needed to confirm.

Establishing the WN convex hull allows us to evaluate the likelihood
of finding other predicted forms of W$_x$N$_{1-x}$. We find that the
P$_3$Tc structure predicted by Song and Wang,\cite{song10:wn} as
well as the ReP$_4$ structure predicted by Aydin {\em et al.}, are
both well above the convex hull and thus unlikely to form, at least
without special non-equilibrium processing.  We did find that the
hexagonal WN$_2$ structures predicted by Wang {\em et
  al.}\cite{wang09:wn2} are stable, and of comparable enthalpy to
the NbO phase -- at least for calculations using the
generalized-gradient PBE functional and the LDA functional.

The problem of determining the convex hull in WN is exacerbated by
the fact that the ground state and other low-energy structures of
N$_2$ are van der Waals solids. As shown in Table~\ref{tab:N2},
neither the LDA nor the PBE functionals adequately describe the
ground state $\alpha$N$_2$. Both correctly describe the bond length
of the N-N dimer, but LDA substantially underestimates the
equilibrium lattice constant while PBE drastically overestimates
it. Since molecular nitrogen is van der Waals bound, we also looked
at the system using the vdW-DF2 functional proposed by Dion {\em et
  al.}\cite{dion04:vdwdft}, which is readily implementable in
VASP.\cite{klimes10:vdwdft,klimes11:vdWtest} This produces a much
better, although still imperfect, lattice constant for
$\alpha$N$_2$.

While it is not entirely clear that the vdW-DF2 functional can be
used to study dense bulk systems, we have applied it to this system
across all compositions. As seen in Fig.~\ref{fig:vdwhull}, this
produces little change on the tungsten-rich side of the phase
diagram, but a dramatic change on the nitrogen-rich side. The
vdW-DF2 functional favors the cubic cI36 structure
(Fig.~\ref{fig:cI36} and Table~\ref{tab:cI36} for WN$_2$, making it
competitive in enthalpy/atom with the NbO structure. The hexagonal
WN$_2$ structures, which are the ground state structures in the LDA
and PBE, move up significantly. This is consistent with the limited
amount of experimental data we have: no bulk hexagonal WN$_2$ phase
has been seen. Meanwhile the ground state NaCl-like structures, with
vacancies, have been seen experimentally.

Much work remains to be done on this system. We have essentially
ignored the hexagonal $\delta$ phases,\cite{khitrova62:WN} which are
seen experimentally in thin films. This requires yet another
exhausting search for vacancy patterns in supercells, this time
starting from hexagonal and rhombohedral unit cells. Given the
relatively low energy position of the $\delta_H^I$ phase, especially
in the LDA calculation, a proper study of vacancy positions is
likely to change the right-hand side of the convex hull.

We also
need to study the effects of pressure. Does the application of
pressure appreciably change the phase diagram, especially on the
right-hand side? Given the extremely open nature of the cI36
structure, even modest pressure may trigger a first-order phase
transition to another structure. We also need to perform more
calculations to establish the validity of the vdW-DF2 functional
when used in bulk systems. We will pursue these calculations in the
future.

In conclusion, we have used high-throughput density functional
calculations to study the tungsten-nitrogen phase diagram, both with
and without van der Waals forces. Our calculations agree with
experiment in that the dominant structure of W-N is the NaCl structure
with vacancies, and show that this structure can persist over a
large range of compositions. It is unlikely that many of the
previously predicted nitrogen-rich phases will ever be seen
experimentally, as they are well above the ground state hull. The
only exceptions may be the hexagonal WN$_2$ phases predicted by Wang
{\em et al.}\cite{wang09:wn2} These are on the ground state hull
when we use the LDA and PBE functionals, but not when we include van
der Waals forces. In addition, the bulk and shear moduli of the
predicted cubic ground state has rather large bulk and shear moduli,
comparable with any of the previously predicted ultra-incompressible
structures.\cite{wang09:wn2,song10:wn,aydina12:XN4}

\section{Acknowledgments}
\begin{acknowledgments}
Most computational work was done at the US Naval Academy and at the
ERDC and AFRL High Performance Computer Centers of the
U.S. Department of Defense. We thank Fulton Supercomputing
Laboratory and the Cray Corporation for additional computational
support. M.J.M. is supported by the Office of Naval
Research. D.F. was supported by the Naval Research Laboratory-U.S.
Naval Academy Cooperative Program for Scientific
Interchange. C.D. was supported by the ``Simulation of Materials
Under Pressure'' Internship at the Naval Research
Laboratory. S.C. acknowledges support from DOD-ONR (Grants
No. N00014-13-1-0635, No. N00014-11-1 0136, and
No. N00014-09-1-0921). G.L.W.H. is grateful for support from the
National Science Foundation, Grant No.DMR-0908753. The authors thank
Y. Ciftci for providing us with the starting coordinates for the
ReP$_4$ structure used in Ref.~\onlinecite{aydina12:XN4}, and also
thank Dr. Ohad Levy and Dr. Allison Stelling for useful comments.
\end{acknowledgments}
l

\end{document}